\documentclass{article}

\usepackage[utf8]{inputenc} % allow utf-8 input
\usepackage[T1]{fontenc}    % use 8-bit T1 fonts

\usepackage{amsmath, amssymb, amsthm, mathtools}
\usepackage{amsfonts}       % blackboard math symbols
\usepackage{nicefrac}       % compact symbols for 1/2, etc.
\usepackage{relsize}

\usepackage{graphicx}       % graphics
\graphicspath{{media/}}     % organize your images and other figures under media/ folder
\usepackage{subcaption}     % For side-by-side figures with captions (subfig is incompatible)
\usepackage{tikz}
\usetikzlibrary{arrows.meta, positioning}

\usepackage{algorithmic}
\usepackage{algorithm}

\usepackage{booktabs}       % professional-quality tables
\usepackage[table,x11names]{xcolor}
\usepackage{arydshln}
\usepackage{float}

\usepackage{geometry}       % For adjusting margins
\usepackage{microtype}      % microtypography
\usepackage{fancyhdr}       % header
\usepackage{authblk}        % For author and affiliation formatting
\usepackage{hyperref}       % hyperlinks
\usepackage{url}            % simple URL typesetting
\usepackage{lipsum}         % for placeholder text
\usepackage{enumitem}
\title{Quantifying the Risk of Long-Term Chikungunya Persistence in Miami-Dade County}
\author{Antonio Gondim}
\author{Leonardo Schultz}
\author{Xi Huo}
\author{Shigui Ruan}
\affil{Department of Mathematics, University of Miami, Coral Gables, FL 33156, USA\thanks{Research was partially supported by National Science Foundation grant (DMS-2424605).}}
\date{October, 2025}

\begin{document}
\maketitle

\begin{abstract}
Chikungunya virus (CHIKV) is a mosquito-borne arbovirus with the potential to establish sustained transmission in subtropical regions like Florida, where climatic and ecological conditions support vector proliferation. In this study, we develop a Continuous-Time Markov Chain (CTMC) model to assess the probability of long-term CHIKV establishment in Miami-Dade County following repeated introductions of external infectious individuals. This work aims to identify seasonal windows of heightened endemic risk and evaluates the impact of vector control strategies—specifically, reductions in mosquito biting rates and carrying capacity—on mitigating the likelihood of persistent transmission. These results generate insights into the dynamics of CHIKV and inform targeted interventions to prevent its transition from minor sporadic outbreaks to endemic circulation.
\end{abstract}

\section{Introduction}

Chikungunya virus (CHIKV) is an arthropod-borne alphavirus that has emerged as a significant public health threat due to its rapid spread and reduced herd immunity. First identified in Africa in the 1950s, CHIKV has since expanded to multiple continents, including Asia, Europe, and the Americas \cite{Vazquez2023}. The virus is primarily transmitted by \textit{Aedes aegypti} and \textit{Aedes albopictus}, two mosquito species that thrive in urban environments. Clinically, CHIKV infection is characterized by acute fever, rash, and severe polyarthritis, that can potentially lead to long-term morbidity, eventually manifesting symptoms that can endure for years.

In the Americas, particularly in Central and South America, CHIKV has shown a rapid geographic expansion following its introduction in 2013 on the Caribbean island of St. Martin. Since then, the virus has been detected in over 40 countries, including Brazil, where it has become endemic. Phylogenetic studies suggest that the dominant strain in Brazil belongs to the East/Central/South African (ECSA) lineage, which has demonstrated increased adaptability to local mosquito populations, thereby enhancing the potential for sustained transmission in previously non-endemic regions \cite{Akhi2024}. This adaptability raises concerns about CHIKV outbreaks in the southern United States, specially in south Florida, where environmental conditions, such as weather and precipitation indices, are conducive to vector survival and propagation.

\begin{figure}[htbp]
    \centering
    \begin{subfigure}[c]{0.45\textwidth}
        \centering
        \includegraphics[width=\textwidth]{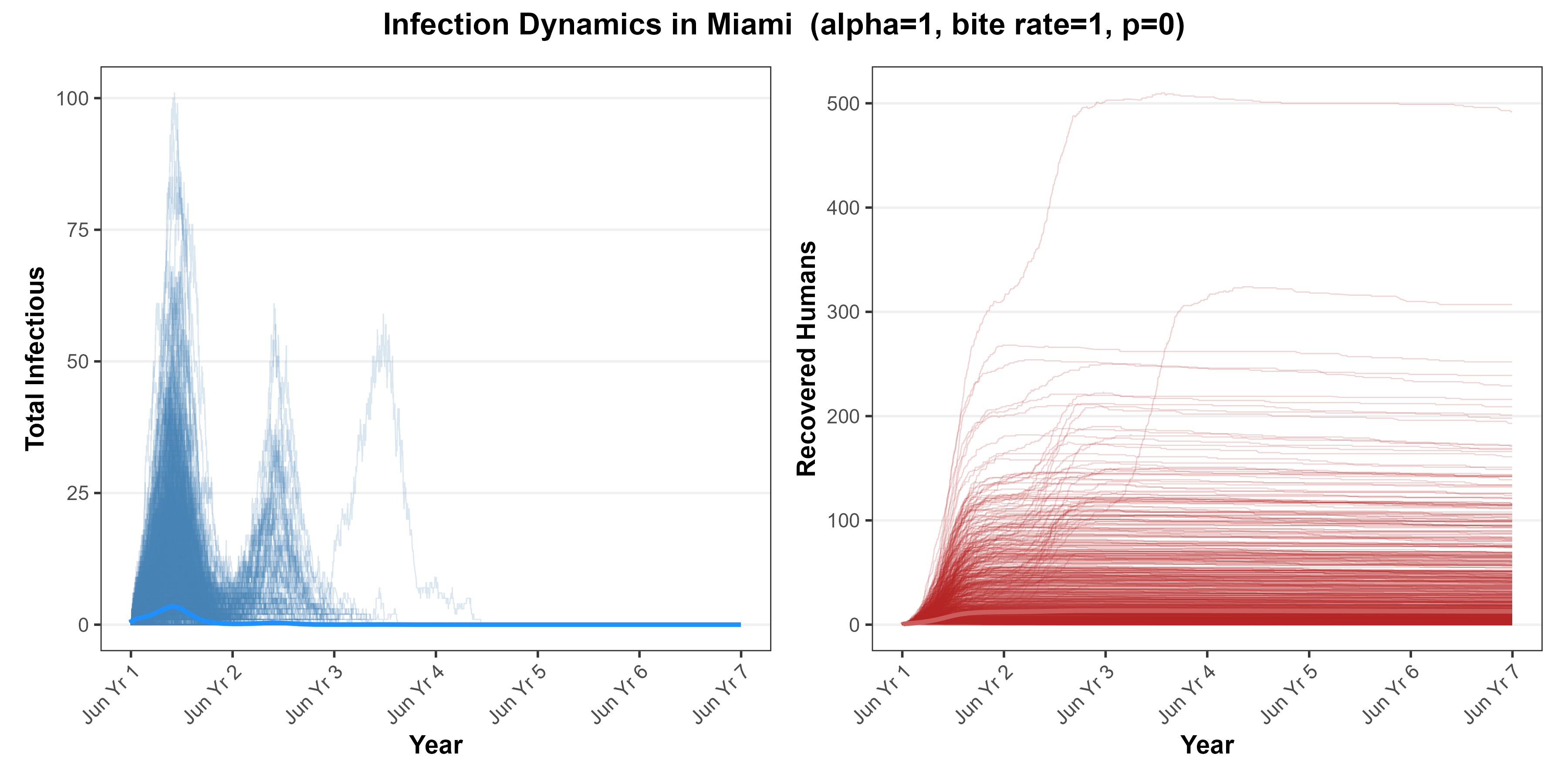}
        \caption{ p=0}
        \label{}
    \end{subfigure}
    \hfill
    \begin{subfigure}[c]{0.45\textwidth}
        \centering
        \includegraphics[width=\textwidth]{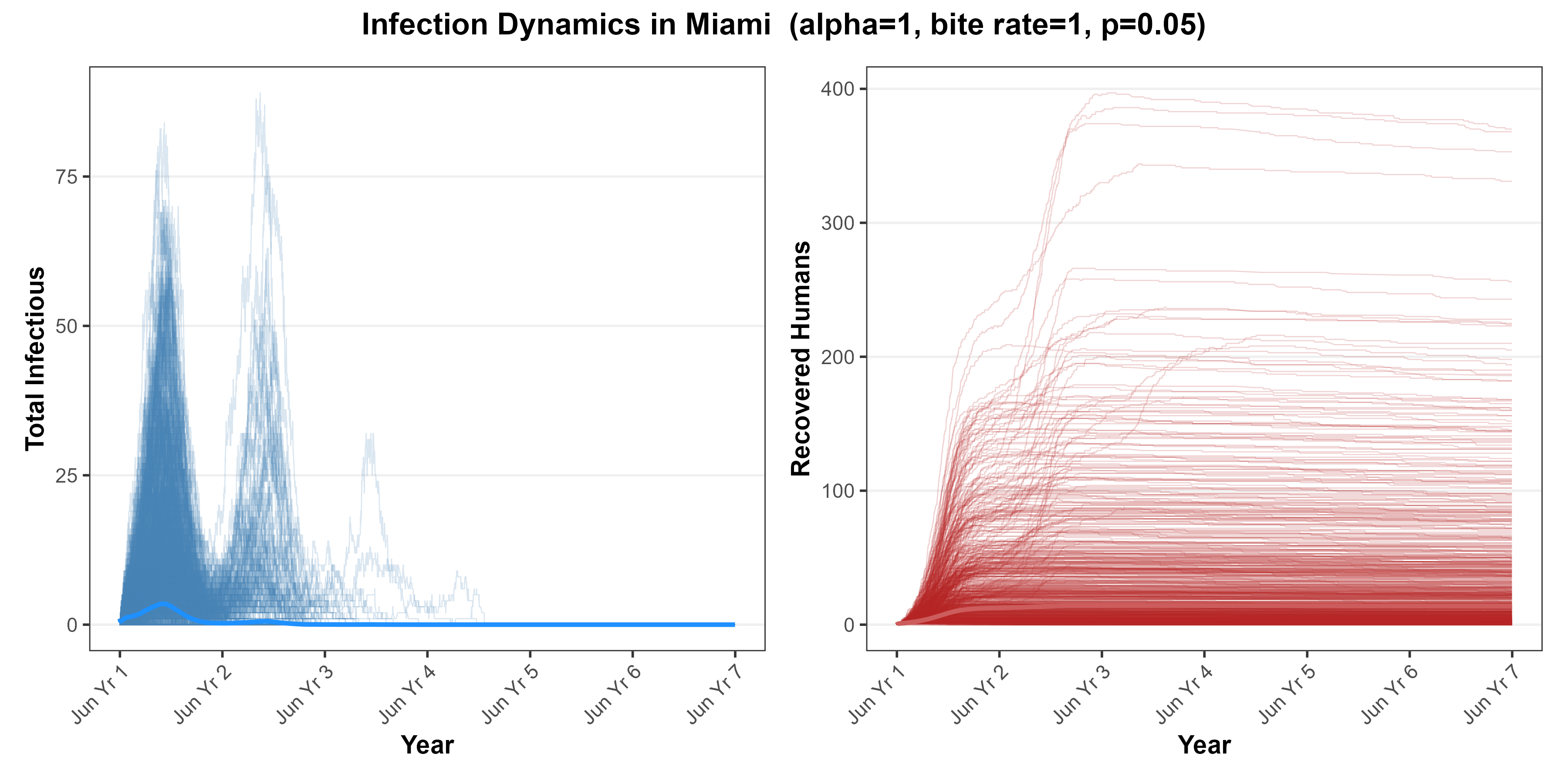}
        \caption{ p=0.05}
        \label{}
    \end{subfigure}
    \vspace{0.05cm}
    \begin{subfigure}[b]{0.45\textwidth}
        \centering
        \includegraphics[width=\textwidth]{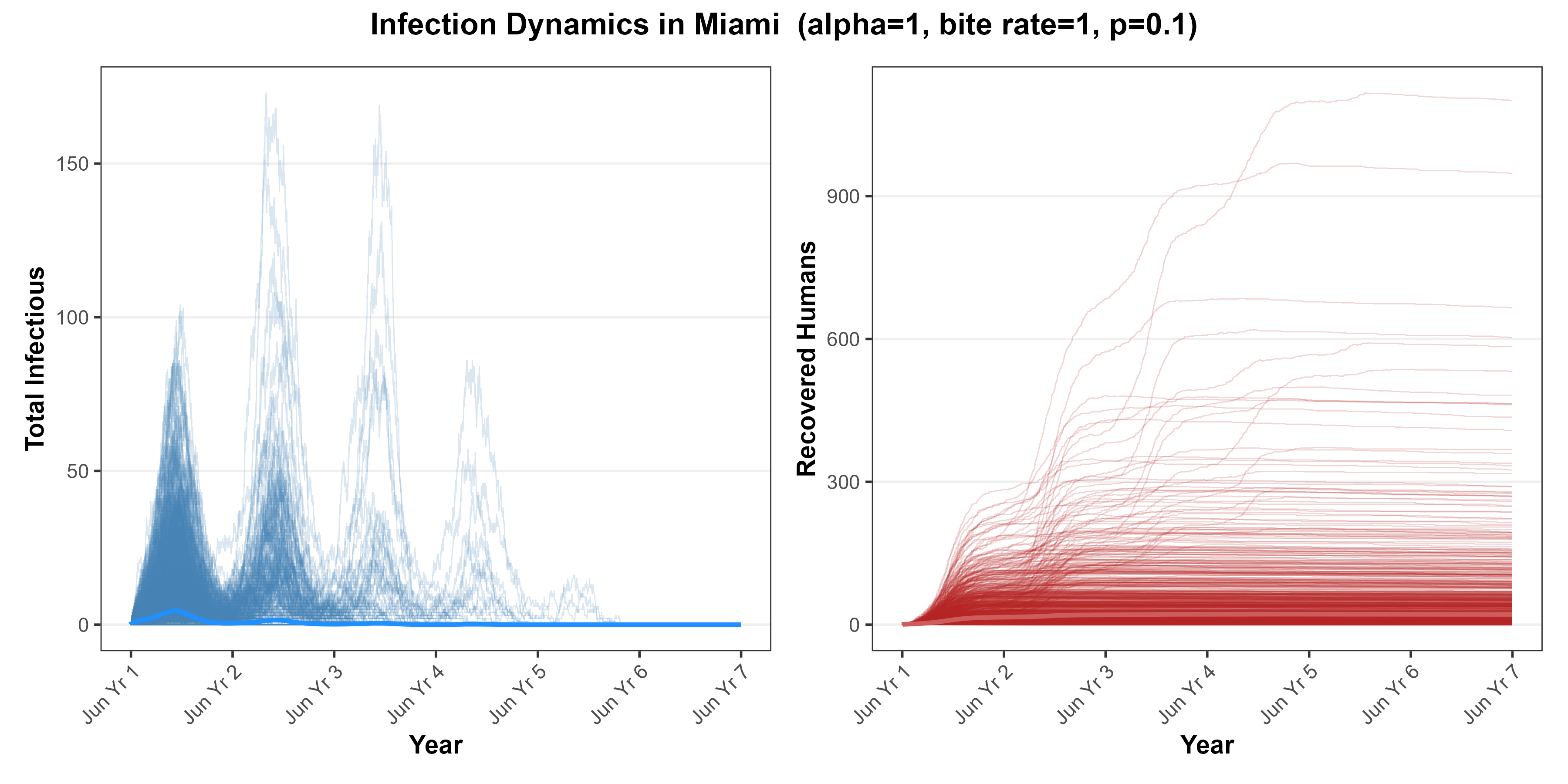}
        \caption{p=0.1 }
        \label{fig:plot3}
    \end{subfigure}
    \hfill
    \begin{subfigure}[b]{0.45\textwidth}
        \centering
        \includegraphics[width=\textwidth]{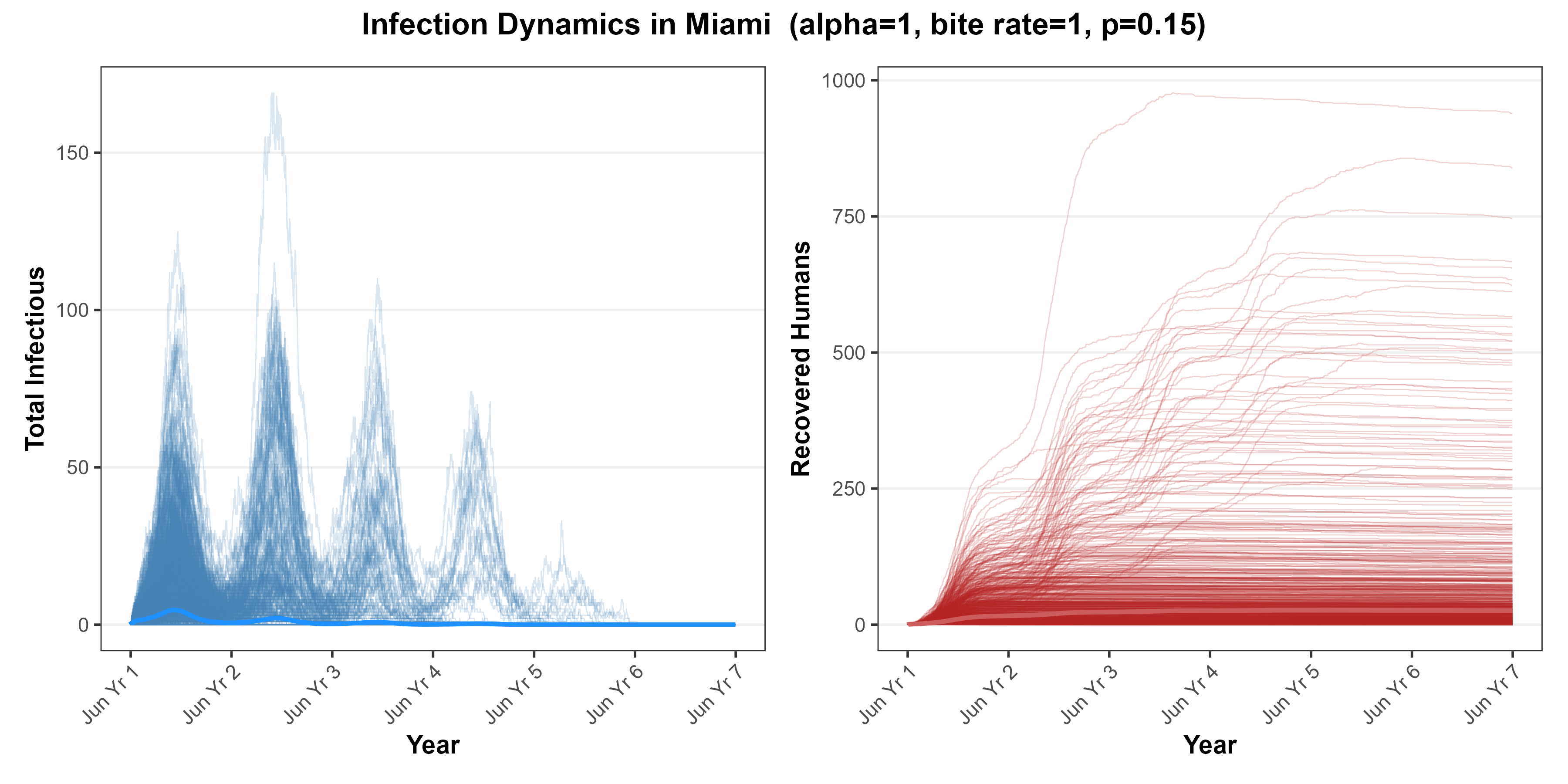}
       \caption{p=0.15 }
        \label{}
    \end{subfigure}
    \caption{Infectious dynamics for different values of vertical probability $p \in [0,0.15]$, given an introduction of one infectious individual on June $29^{th}$ which is the initial time on the plot. In total, 2000 trajectories are plotted and for each figure the solid lines represent the mean.}
    \label{fig:five_plots}
\end{figure}

Florida, and particularly Miami-Dade County, represents a high-risk region for CHIKV transmission due to its warm climate, high human population density, and mainly due to its extensive travel connections with endemic regions, such as Brazil, the Caribbean and Southeast Asia \cite{CDCChikungunya2014}. In 2014, Florida has primarily reported travel-associated CHIKV and 12 local acquired cases \cite{FloridaHealthChikungunya2014}. The presence of competent Aedes mosquito vectors, combined with increasing global temperatures, changing precipitation patterns, and urban expansion, may contribute to the establishment of local transmission. These factors emphasize the need for mathematical modeling approaches to assess outbreak potential and inform public health interventions. Also, in this study, we consider the probability of vertical transmission, as demonstrated experimentally in \cite{Honorio2019}, where, as shown in Figure \ref{fig:five_plots0} and Figure \ref{fig:five_plots}, we are able to analyze its relevance on quantifying the long-term of disease persistence.

Mathematical modeling plays a crucial role in understanding the transmission dynamics of vector-borne diseases, such as CHIKV. Deterministic models, such as \cite{gondim2025, Schultz}, typically formulated using differential equations, are largely employed to estimate key model parameters in disease modeling. However, deterministic models fail to account for stochastic fluctuations, particularly in the early stages of an outbreak or when environmental variability significantly influences mosquito populations \cite{Hridoy2025}, \cite{ALLEN2017128}, \cite{Allen2021}. To address these limitations, stochastic models, such as Continuous-Time Markov Chains (CTMCs) and Multitype Branching Processes, provide a more refined representation of outbreak dynamics by incorporating probabilistic disease extinction scenarios \cite{Bacaer2014, Neal2025}.

In the present study, we construct a CTMC model to quantify the probability of CHIKV outbreak in Miami-Dade County. The model incorporates time-dependent parameters to address the influence of seasonal effects due to variations in temperature. In Miami-Dade, as observed in \cite{CHEN2023106837}, the temperature and rainfall are significantly correlated, therefore it is sufficient to let the time dependent parameters of this model be temperature dependent uniquely, in order to address seasonal variability. These seasonal changes significantly impact the life cycle of mosquitoes, which in turn affects the dynamics of the disease, i.e, the number of infections. By accounting for these factors, the model can more accurately predict the spread and intensity of mosquito-borne diseases throughout different times of the year. Additionally, to simulate external periodical introductions of infectious humans, we will be considering an external infectious compartment for humans through a periodical introduction function $f: \mathbb{R_+}\rightarrow \mathbb{R_+}.$

By employing a Branching Process Approximation (BPA), we derive the probability of disease extinction, by solving a periodical system of ODEs, first given that a fixed number of introductions of infectious compartments is initially introduced in a susceptible population; then, considering periodical external introductions of infectious humans into a totally susceptible population. Numerical simulations are conducted to validate our theoretical predictions, along with sensitivity analyses of relevant parameters, in order to explore their influence on the probability of disease extinction. The findings of this study contribute to a more comprehensive understanding of CHIKV transmission risk through insertion of external infected individuals into a totally susceptible population in Miami-Dade, and provide insights for targeted public health interventions aimed at mitigating outbreak potential and specially disease persistency in Miami-Dade County.

The rest of the paper is organized as follows: in Section 2, the deterministic model is constructed. In Section 3, the CTMC models and Branching Process Approximations with and without periodic external introductions are presented. In Section 4, numerical simulations of the CTMC model and probability of disease extinction are presented. A brief discussion is given in Section 5.

\section{The Deterministic Model Formulation}
\label{sec:deterministic}

In this section, we develop a deterministic compartmental model for the transmission dynamics of Chikungunya virus (CHIKV) between mosquitoes and humans. The model consists of three main interacting components: the human population, the adult mosquito population, and the mosquito larvae population. All variables and parameters of the model are shown in Table \ref{table1}. We assume that all time dependent parameters are positive $w-$periodic continuous functions, existing $w>0,$ except for $f:\mathbb{R_+}\rightarrow\mathbb{R_+}$ which is considered to be a nonnegative periodic piece-wise continuous function.
We consider the local population of humans to be constant, i.e, let for every $t\geq 0,$ $N_{h}:=S_h (t) +E_h (t) + I_h (t) +R_h (t),$ which is a positive constant, according to System (\ref{ode1}), where $S_h (t) ,E_h (t) , I_h (t) ,R_h (t)$ denote the number of susceptible, exposed, infectious and recovered individuals at time $t\geq 0.$

The model is described by the following System (\ref{ode1}):

\begin{align}\label{ode1}
\frac{dS_h}{dt} &= \mu_h N_h - b_m(t) \beta_{mh} \frac{I_m S_h}{N_h} - \mu_h S_h  \\
\frac{dE_h}{dt} &= b_m(t) \beta_{mh} \frac{I_m S_h}{N_h} - v_h E_h - \mu_h E_h \notag\\
\frac{dI_h}{dt} &= v_h E_h - \eta_h I_h - \mu_h I_h \notag\\
\frac{dI_{ext}}{dt} &=f(t) - \mu I_{ext} \notag
\\
\frac{dR_h}{dt} &= \eta_h I_h - \mu_h R_h \notag\\
\frac{dS_m}{dt} &= \eta_l(t) L_S - b_m(t) \beta_{hm} \frac{I_h S_m}{N_h} - \mu_m(t) S_m \notag \\
\frac{dE_m}{dt} &= b_m(t) \beta_{hm} \frac{(I_h + I_{ext}) S_m}{N_h} - \eta_m E_m - \mu_m(t) E_m \notag \\
\frac{dI_m}{dt} &= \eta_m E_m - \mu_m(t) I_m + \eta_l(t) L_I \notag \\
\frac{dL_I}{dt} &= \eta_b(t) \left(1 - \frac{L_I + L_S}{K} \right) p I_m - \eta_l(t) L_I - \mu_l(t) L_I \notag \\
\frac{dL_S}{dt} &= \eta_b(t) \left(1 - \frac{L_S + L_I}{K} \right) (S_m + E_m + (1 - p) I_m) - \eta_l(t) L_S - \mu_l(t) L_S.\notag
\end{align}

Table \ref{table1} addresses all the state variables and parameters in the model.

\begin{table}
    \centering
\label{tabelinha}
\caption{Variables and Parameters Used in the Model for Human and Mosquito Populations}

\begin{tabular}{|c|c|}
\hline\label{table1}
$S_{h}$ & Susceptible humans  \\
$E_{h}$ & Exposed humans \\
$I_{h}$ & Infected humans\\
$I_{ext}$ & Introductions of infectious humans\\
$R_{h}$ & Recovered humans\\
$S_{m}$ & Susceptible mosquitoes\\
$E_{m}$ & Exposed mosquitoes\\
$I_{m}$ & Infected mosquitoes\\
$L_{s}$ & Susceptible larvae\\
$L_{I}$ & Infected larvae\\
$R_h$ & Human population recruitment rate\\
$\mu_{h}$ & Human natural death rate \\
$\mu_{m}(t)$ & Mosquito natural death rate\\
$b_{m}(t)$ & Per capita mosquito biting rate\\
$\beta_{mh}$ & Human-mosquito transmission probability \\
$\beta_{hm}$ & Mosquito-human transmission probability \\
$\eta_{h}$ & Rate of recovery for humans \\
$\mu_{b}(t)$ & Larvae birth rate\\
$\eta_l(t)$ & Larvae development rate\\
$\eta_h$ & Human rate of progression from exposed to infectious\\
$p$ & Vertical transmission probability \\%https://www.ncbi.nlm.nih.gov/pmc/articles/PMC6520672/
$K$ & Carrying capacity\\
$\mu_{l}(t)$ & Larvae death rate\\
\hline
\end{tabular}

\end{table}

To apply the Branching Process Approximation, we linearize the System \ref{ode1} at the Disease-Free Equilibrium (DFE), where there are no infected humans or mosquitoes present in the population. We choose to linearize the system around the following time-dependent DFE:

\begin{equation}\label{DFE}
(N_h, 0,0, 0, 0, S_m^*(t), 0, 0, 0, L_S^*(t)),
\end{equation}We emphasize that in this equilibrium, the mosquito and larvae populations, \( S_m^*(t) \) and \( L_S^*(t) \), are positive $w-$periodical trajectories, given by \begin{align*}L_S^{*}(t) &= K(t)\left(1 - \frac{\mu_m(t)\left(\eta_l(t) + \mu_l(t)\right)}{\eta_l(t)\eta_b(t)}\right)
\\
S_m^*(t) &=\frac{\eta_l(t)L_S^*(t)}{\mu_m(t)}.
\end{align*} This choice reflects the scenario in Miami-Dade county, where mosquito populations are consistently present due to favorable environmental conditions. 
%\begin{align} 
%\frac{dS_h}{dt} &= \mu_h N_h - b_m(t) b_{mh} I_m - \mu_h S_h \\
%\frac{dE_h}{dt} &= b_m(t) b_{mh} I_m - v_h E_h - \mu_h E_h \notag \\
%\frac{dI}{dt} &= \notag v_h E_h - \eta_h I_h - \mu_h I_h \\
%\frac{dR_h}{dt} &= \eta_h I - \mu_h R_h \notag\\
%\frac{dS_m}{dt} &= \eta_L(t) L_S - b_m(t) b_{hm} \frac{I S_m^*(t)}{N_h} - \mu_m(t) S_m \notag\\
%\frac{dE_m}{dt} &= b_m(t) b_{hm} \frac{I S_m^*(t)}{N_h} - \eta_m E_m - \mu_m(t) E_m \notag \\
%\frac{dI_m}{dt} &= \eta_m E_m - \mu_m(t) I_m + \eta_L(t) L_I \notag \\
%\frac{dL_I}{dt} &= \eta_b(t) \left(1 - \frac{L_S^*(t)}{K(t)}\right) p I_m - \eta_L(t) L_I - \mu_L(t) L_I \notag \\
%\frac{dL_S}{dt} &= \notag \eta_b(t) \left(1 - \frac{L_S^*(t)}{K(t)}\right) \left(S_m(t) + E_m + (1 - p) I_m\right)  \\
%&\quad - \eta_b(t) \left(\frac{L_S(t) + L_I (t)}{K(t)}\right) S_m^*(t) - \eta_L(t) L_S - \mu_L(t) L_S \notag
%\end{align}

\section{Stochastic Model: CTMC and Branching Process Approximation}
\label{sec:stochastic}
\subsection{CTMC Model and Branching Process Approximation without periodical external introductions}\label{3.1}

When the density of infected mosquitoes or the number of infectious human individuals introduced into a population is sufficiently small, deterministic models may fail to capture the initiation of an outbreak. Specifically, if a single infectious mosquito or human is introduced into a fully susceptible population, there is a probability that the infection will die out before spreading to a sufficient number of susceptible hosts. This could occur if an infectious human recovers or an infected mosquito dies before transmitting the virus effectively. Similarly, if only a few infected mosquitoes are present in the environment, they may die before biting enough susceptible humans to cause a major outbreak.

A time-nonhomogeneous Markov process, where time is sampled from an exponential distribution and the state variables are discrete, can be formulated based on the transition rates from system (\ref{ode1}). The state variables in this model include the numbers of susceptible, exposed, infectious and recovered humans, as well as the numbers of susceptible, exposed, and infectious mosquitoes, along with larval compartments representing immature mosquito stages. 
In the section, we will ultimately consider no introduction of infectious humans, i.e, let $f(t)=\mu=0,\ \forall t\geq 0.$

The random variable is represented as:
\[
X(t) = (S_h(t), E_h(t), I_h(t), R_h(t), S_m(t), E_m(t), I_m(t), L_I(t), L_S(t)),
\] where $\forall t \geq 0,$ $X(t) \in \mathbb{Z}_{+}^{9}$.

The transition probabilities for the CTMC model are given in Table 2.

\begin{table}[h]
        %\centering
        \begin{tabular}{|p{5.5cm}|c|p{6.5cm}|}
    %\begin{tabular}{|l|c|c|}
        \hline
        Description & Transition & Probability Rate \\
        \hline
        Recruitment of susceptible humans & \( S_h \to S_h + 1 \) & \( \mu_h N_h \) \\
        Death of susceptible humans & \( S_h \to S_h - 1 \) & \( \mu_h S_h \) \\
        Human infection via mosquito bite & \( S_h \to S_h - 1, E_h \to E_h + 1 \) & \( b_m(t) \beta_{mh} \frac{I_m S_h}{N_h} \) \\
        Progression from exposed to infectious humans & \( E_h \to E_h - 1, I_h \to I_h + 1 \) & \(  v_h E_h \) \\
        
        Recovery of infectious humans & \( I_h \to I_h - 1, R_h \to R_h + 1 \) & \( \eta_h I_h \) \\
        Death of exposed humans & \( E_h \to E_h - 1 \) & \( \mu_h E_h \) \\
        Death of infectious humans & \( I_h \to I_h - 1 \) & \( \mu_h I_h \) \\
        Death of recovered humans & \( R_h \to R_h - 1 \) & \( \mu_h R_h \) \\
        Recruitment of susceptible mosquitoes &\( S_m \to S_m + 1 \) & \( \eta_L(t) L_S \) \\
        Death of susceptible mosquitoes & \( S_m \to S_m - 1 \) & \( \mu_m(t) S_m \) \\
        Mosquito infection via biting humans & \( S_m \to S_m - 1, E_m \to E_m + 1 \) & \( b_m(t) \beta_{hm} \frac{I S_m}{N_h} \) \\
        Progression from exposed to infectious mosquitoes & \( E_m \to E_m - 1, I_m \to I_m + 1 \) & \( \eta_m E_m \) \\
        Death of infectious mosquitoes & \( I_m \to I_m - 1 \) & \( \mu_m(t) I_m \) \\
        Recruitment of infected larvae & \( L_I \to L_I + 1 \) & \( \eta_b(t) \left(1 - \frac{L_I + L_S}{K} \right) p I_m \) \\
        Recruitment of susceptible larvae & \( L_S \to L_S + 1 \) & \( \eta_b(t) \left(1 - \frac{L_S + L_I}{K} \right) (S_m + E_m + (1 - p) I_m) \) \\
        Death of infected larvae & \( L_I \to L_I - 1 \) & \( \mu_L(t) L_I \) \\
        Death of susceptible larvae & \( L_S \to L_S - 1 \) & \( \mu_L(t) L_S \) \\
        \hline
    \end{tabular}
    \caption{Transition probabilities for the CTMC}
    \label{tab:transition}
\end{table}

%VOU MUDAR O SEGUINTE: AO INVES DE USAR TABLE 3 P CALCULAR BACKWARD KOLMOGOROV EQ, USO A TABLE DO CTMC, E AI CALCULO, E COLOCO AS HIPOTESES Q A SIMPLIFICAM DE MODO A FICAR UM SISTEMA RESOLVIVEL..DO JEITO COMO TA NAO VAI SER ACEITO PQ TAMO ANALISANDO A DOENCA EM MUITOS DIAS E N FAZ SENTIDO DIZER Q A POPULACAO $S_h$ FICA CONSTANTE, POR EXEMPLO. Q EH O Q TA SE APROXIMANDO NA TABELA 3. A APROXIMACAO Q VOU USAR EH A ESTABELECIDA NA PAG 7 MAS PARA A TABLE MAIOR...OUTRA SOLUCAO EH APENAS PLOTAR P/ OS CASOS ONDE $K_{MULT}\leq 1$, Q DE ACORDO COM OS PLOTS DO LEO, N VAI GERAR UMA POPULACAO DE RECUPERADOS GRANDES, AI SIM POSSO CONTINAUR FAZENDO COMO ESTAH!!!!!!!!!!!!!!!!!!!!!!!!!!!!!!!!!!!!!!!!!!!!!!!!!!!!!!!!!!!!!!!!!!!!!!!!!!!!!!!!!!!!!!!!!!!!!!!!!!!!!!!!!!!!!!!!!!!!!!!!!!!!!!!!!!!!!!!!

To derive a stochastic threshold for Chikungunya transmission, we apply the theory of multitype branching processes to approximate the time-nonhomogeneous stochastic process near the DFE. We assume that the total human population remains constant, \( S_h(t) = N_h \), and focus on the infectious random variables, namely the numbers of infectious and exposed humans \( I, E_h \), respectively and the numbers of infectious larvae, infectious mosquitoes \( L_I, I_m \), respectively. 

We analyze the stochastic dynamics near the disease-free equilibrium to determine the probability of a major outbreak, or equivalently, the probability of disease extinction, when a small number of infectious individuals are introduced. Also, as in \cite{NIPA2021}, \cite{Akhi2024} and \cite{Allen2021}, we assume that the infective variables are independent random variables, which is a realistic assumption, once taking into account that in the early stages of the disease each infectious population is sufficiently small comparatively to its respective susceptible population. 

%The branching process models the infectious compartments as a birth-and-death process, while the susceptible and recovered compartments are not explicitly considered. The process begins with a small number of infectious individuals and evolves according to the rates. If the number of infectious individuals grows exponentially, a major outbreak occurs; otherwise, the epidemic dies out with only a few additional cases. This approximation is valid when the susceptible population is sufficiently large.

The approximation relies on three key assumptions:  
\begin{enumerate}[label=(\roman*)]
    \item Infectious individuals transmit the infection independently.
    \item All infectious individuals have the same transmission and recovery probabilities.
    \item The susceptible population is sufficiently large.
\end{enumerate}
\label{BPA_assumptions} % Label for the entire list

Assumption (i) is reasonable if a small number of infectious individuals is introduced into a large homogeneously-mixed
population, i.e, Assumption (iii). Assumption (ii) is also reasonable in a homogeneously-mixed population with constant transmission and recovery rates.

Under these conditions, the Branching Process captures the early-stage dynamics of the CTMC model. For this, we should first notice that when all infective variables reach the value zero, and when no external introductions of infectious humans are considered, then the system is in absorption state. We define this phenomena as being the extinction of the disease. Due to the stochasticity of the model, it can be reached at a finite time, differently from deterministic models.

The following table addresses the probability rates for the Branching Process Approximation, which are basically a linearization over the DFE of the rates taken from Table \ref{tab:transition}, i.e, the table represents the rates seen in system (\ref{sec:deterministic}).

\begin{table}[h]
    \centering
    \begin{tabular}{|l|c|c|}
        \hline
        Description & Transition & Probability Rate \\
        \hline
        Human infection via mosquito bite & \( E_h \to E_h + 1 \) & \( b_m(t) \beta_{mh} I_m \) \\
        Progression from exposed to infectious humans & \( E_h \to E_h - 1, I_h \to I_h + 1 \) & \(  v_h E_h \) \\
        Recovery of infectious humans & \( I_h \to I_h - 1 \) & \( \eta_h I_h \) \\
        Death of exposed humans & \( E_h \to E_h - 1 \) & \( \mu_h I_h \) \\
        Death of infectious humans & \( I_h \to I_h - 1 \) & \( \mu_h I_h \) \\
        Mosquito infection via biting humans & \( E_m \to E_m + 1 \) & \( b_m(t) \beta_{hm} \frac{I S_m^{*}(t)}{N_h} \) \\
        Progression from exposed to infectious mosquitoes & \( E_m \to E_m - 1, I_m \to I_m + 1 \) & \( \eta_m E_m \) \\
        Death of infectious mosquitoes & \( I_m \to I_m - 1 \) & \( \mu_m(t) I_m \) \\
        Recruitment of infected larvae & \( L_I \to L_I + 1 \) & \( \eta_b(t) \left(1 - \frac{L_S^*(t)}{K} \right) p I_m \) \\ 
        Death of infected larvae & \( L_I \to L_I - 1 \) & \( \mu_L(t) L_I \) \\
        \hline
    \end{tabular}
    \caption{Transition probabilities for the Branching Process Approximation}
    \label{tab:BPA}
\end{table}

For the Branching Process, let $t >0$, $\tau\in [0,t)$, and as the process is time-nonhomogeneous we define the time dependent transition probability from state $(i_1,i_2,i_3,i_4,i_5)\in \mathbb{Z}^{5}_{+}$ to $(j_1,j_2,j_3,j_4,j_5)\in \mathbb{Z}^{5}_{+}$, which keep track of the initial infectious variables introduction time $\tau>0$ and a final fixed time $t>\tau$, as \begin{align*} 
\mathbb{P}_{(i_1,i_2,i_3,i_4,i_5), (j_1,j_2,j_3,j_4,j_5)}(\tau, t) &= \mathbb{P}((E_{h}(t), I_h(t), E_m (t), I_{m}(t), L_{I}(t))\\&=(i_1,i_2,i_3,i_4,i_5)|(E_{h}(t), I(t), E_m (t), I_{m}(t), L_{I}(t))=(j_1,j_2,j_3,j_4,j_5)).
\end{align*}

Next, we derive the Backward Kolmogorov differential equations. Hence, fix $t>\tau>0$, and let $\Delta \tau>0$ be sufficiently small, so that only one of the events depicted in the Table \ref{tab:BPA} will occur in this time interval. Thus, by the Table \ref{tab:BPA}, we arrive at \begin{align*} \tiny
\mathbb{P}_{(i_1,i_2,i_3,i_4,i_5), (j_1,\dots, j_5)}(\tau - \Delta \tau, t) & = \big[  b_{m}(\tau - \Delta \tau)\beta_{mh}i_4\mathbb{P}_{(i_1 +1  ,i_2,i_3,i_4,i_5), (j_1,\dots, j_5)}(\tau, t)  \\& +  \mu_h i_1\mathbb{P}_{(i_1 -1 ,i_2,i_3,i_4,i_5), (j_1,\dots, j_5)}(\tau, t)  \\& + v_h i_1 \mathbb{P}_{(i_1 -1,i_2 +1,i_3,i_4,i_5), (j_1,\dots, j_5)}(\tau, t)  \\& + (\eta_h + \mu_h) i_2 \mathbb{P}_{(i_1 ,i_2 -1,i_3,i_4,i_5), (j_1,\dots, j_5)}(\tau, t) \\&+ b_{m}(\tau - \Delta \tau ) \beta_{hm}i_2 \frac{S_{m}^{*}(\tau - \Delta \tau)}{N_h}\mathbb{P}_{(i_1 ,i_2,i_3+1,i_4,i_5), (j_1,\dots, j_5)}(\tau, t)  \\& + \mu_{m}(\tau - \Delta \tau) i_3 \mathbb{P}_{(i_1 ,i_2,i_3 -1,i_4,i_5), (j_1,\dots, j_5)}(\tau, t) 
\\& + \eta_m(\tau - \Delta \tau) i_3 \mathbb{P}_{(i_1,i_2,i_3-1,i_4+1,i_5), (j_1,\dots, j_5)}(\tau, t) 
\\&+\mu_m (\tau - \Delta \tau) i_4 \mathbb{P}_{(i_1,i_2,i_3,i_4 - 1,i_5), (j_1,\dots, j_5)}(\tau, t) 
\\&+p i_4\eta_b (\tau-\Delta \tau)\big( 1 - \frac{L_{S}^{*}(\tau - \Delta \tau)}{K}  \big)\mathbb{P}_{(i_1,i_2,i_3,i_4,i_5+1), (j_1,\dots, j_5)}(\tau, t)
\\& + \eta_{l}(\tau - \Delta \tau)i_5 \mathbb{P}_{(i_1,i_2,i_3,i_4 +1,i_5 -1), (j_1,\dots, j_5)}(\tau, t) 
\\&+\mu_{l}(\tau - \Delta \tau)i_5 \mathbb{P}_{(i_1,i_2,i_3,i_4,i_5 -1), (j_1,\dots, j_5)}(\tau, t)\big]\Delta \tau \\&+ \{1 - \big[b_{m}(\tau - \Delta \tau)\beta_{mh}i_5+\mu_h i_1 + v_h i_1+ (\eta_h + \mu_h) i_2 \\&+ b_{m}(\tau - \Delta \tau ) \beta_{hm}i_2 \frac{S_{m}^{*}(\tau - \Delta \tau)}{N_h}+ \mu_{m}(\tau - \Delta \tau) i_3
 + \eta_m(\tau - \Delta \tau)+\mu_m (\tau - \Delta \tau) i_4\\&+p i_4\eta_b (\tau-\Delta \tau)\big( 1 - \frac{L_{S}^{*}(\tau - \Delta \tau)}{K}  \big)\\&+ \eta_{l}(\tau - \Delta \tau)i_5 +\mu_{l}(\tau - \Delta \tau)i_5\big]\Delta \tau \}\mathbb{P}_{(i_1,i_2,i_3,i_4,i_5), (j_1,\dots, j_5)}(\tau, t) +o(\Delta \tau).
\end{align*}
Thus, defining $P^{i}(\tau, t):=\mathbb{P}_{(i_1,i_2,i_3,i_4,i_5), (0,\dots, 0)}(\tau, t) $, subtracting by $P^i(\tau, t)$ and letting $\Delta \tau \rightarrow 0$, we obtain
\begin{align*}
-\frac{d}{d\tau} P^i(\tau, t) & = b_{m}(\tau )\beta_{mh}i_4(\mathbb{P}_{(i_1 +1  ,i_2,i_3,i_4,i_5), (0,\dots, 0)}(\tau, t)-P^{i}(\tau, t))  \\& +  \mu_h i_1(\mathbb{P}_{(i_1 -1 ,i_2,i_3,i_4,i_5), (0,\dots, 0)}(\tau, t)-P^{i}(\tau, t))  \\& + v_h i_1 (\mathbb{P}_{(i_1 -1,i_2 +1,i_3,i_4,i_5), (0,\dots, 0)}(\tau, t)-P^{i}(\tau, t))  \\& + (\eta_h + \mu_h) i_2 (\mathbb{P}_{(i_1 ,i_2 -1,i_3,i_4,i_5), (0,\dots, 0)}(\tau, t)-P^{i}(\tau, t)) \\&+ b_{m}(\tau  ) \beta_{hm}i_2 \frac{S_{m}^{*}(\tau)}{N_h}(\mathbb{P}_{(i_1 ,i_2,i_3+1,i_4,i_5), (0,\dots, 0)}(\tau, t)-P^{i}(\tau, t))  \\& + \mu_{m}(\tau) i_3 (\mathbb{P}_{(i_1 ,i_2,i_3 -1,i_4,i_5), (0,\dots, 0)}(\tau, t)-P^{i}(\tau, t)) 
\\& + \eta_m(\tau) i_3 (\mathbb{P}_{(i_1,i_2,i_3-1,i_4+1,i_5), (0,\dots, 0)}(\tau, t)-P^{i}(\tau, t)) 
\\&+\mu_m (\tau ) i_4 (\mathbb{P}_{(i_1,i_2,i_3,i_4 - 1,i_5), (0,\dots, 0)}(\tau, t) -P^{i}(\tau, t))
\\&+p i_4\eta_b (\tau)\big( 1 - \frac{L_{S}^{*}(\tau) }{K}  \big)(\mathbb{P}_{(i_1,i_2,i_3,i_4,i_5+1), (0,\dots, 0)}(\tau, t)-P^{i}(\tau, t))
\\& + \eta_{l}(\tau )i_5 (\mathbb{P}_{(i_1,i_2,i_3,i_4 +1,i_5 -1), (0,\dots, 0)}(\tau, t) -P^{i}(\tau, t))
\\&+\mu_{l}(\tau)i_5 (\mathbb{P}_{(i_1,i_2,i_3,i_4,i_5 -1), (0,\dots, 0)}(\tau, t)-P^i(\tau, t)).
\end{align*}Define 

\begin{align*}  P_1(\tau, t) &=  \mathbb{P}_{(1,0,0,0,0), (0,\dots, 0)}(\tau, t), \\P_2(\tau, t) &=  \mathbb{P}_{(0,1,0,0,0), (0,\dots, 0)}(\tau, t), \\ P_3(\tau, t) &=  \mathbb{P}_{(0,0,1,0,0), (0,\dots, 0)}(\tau, t),\\ P_4(\tau, t) &=  \mathbb{P}_{(0,0,0,1,0), (0,\dots, 0)}(\tau, t),\\ P_5(\tau, t) &=  \mathbb{P}_{(0,0,0,0,1), (0,\dots, 0)}(\tau, t). \end{align*}
Now, due to the Branching Process Approximation assumptions (i) through (iii), we assume that \begin{align*}
    \mathbb{P}_{(1,0,0,1,0), (0,\dots, 0)}(\tau, t)&= P_{1}(\tau, t)P_4(\tau, t),\\
    \mathbb{P}_{(0,0,0,1,1), (0,\dots, 0)}(\tau, t)&= P_{5}(\tau, t)P_4(\tau, t),\\
    \mathbb{P}_{(0,1,1,0,0), (0,\dots, 0)}(\tau, t)&= P_{2}(\tau, t)P_3(\tau, t).
\end{align*} 
%QUANDO FOR PLOTAR OS GRAFICOS DAS PROBABILIDADES DE EXTINCAO, PLOTAR MOSTRANDO AS DIFERENTES INTRODUCOES DE ACORDO COM ESSAS HIPOTESES ASSUMIDAS!!!!!!!!!!!!!!!!!!!!!!!!!!!
 
This simplification will allow us to obtain the 5-dimensional system: 

\begin{align*}
\tiny
\frac{dP_1}{d\tau}(t,\tau) &= -\mu_h(1-P_1(\tau,t)) -v_h (P_2 (\tau, t) - P_1(\tau, t))\\
\tiny \frac{dP_2}{d\tau}(t,\tau) &= -(\eta_h + \mu_h)(1-P_2(\tau, t)) - b_{m}(\tau)\beta_{hm}\frac{S_{m}^{*}(\tau)}{N_h}(P_3(\tau, t)P_{2}(\tau, t)-P_{2}(\tau, t))\\
\tiny \frac{dP_3}{d\tau}(t,\tau) &= -\mu_m(\tau)(1-P_3(\tau, t)) - \eta_m(\tau)(P_4(\tau, t)-P_3(\tau,t))\\
\tiny \frac{dP_4}{d\tau}(t,\tau) &=   -b_m(\tau)\beta_{mh}(P_{1}(\tau, t)P_{4}(\tau, t)-P_{4}(\tau,t)) -  \mu_m (\tau)(1-P_4(\tau, t))-p \eta_b(\tau)\bigg{(} 1- \frac{L_{S}^{*}(\tau)}{K}  \bigg{)}P_4(\tau, t)(P_5(\tau, t) - 1)\\
\tiny \frac{dP_5}{d\tau}(t,\tau) &= -\eta_{l}(\tau)(P_4(\tau, t)-P_5(\tau, t))-\mu_{l}(\tau)(1-P_5(\tau, t)),\end{align*}for $t>0$ fixed and $\tau \in [0,t)$.

Now let $s=t - \tau$ and define $Q_k (s) = P_k (t-s,t),$ then we arrive in the following system
\begin{align}\label{ext_prob}
\frac{dQ_1}{ds} &= -\mu_h(1-Q_1) -v_h (Q_2  - Q_1)\\
\notag\tiny \frac{dQ_2}{ds} &= -(\eta_h + \mu_h)(1-Q_2) - b_{m}(t-s)\beta_{hm}\frac{S_{m}^{*}(t-s)}{N_h}(Q_3 Q_{2}-Q_{2})\\
\notag\tiny \frac{dQ_3}{ds} &= -\mu_m(t-s)(1-Q_3) - \eta_m(t-s)(Q_4-Q_3)\\
\notag\tiny \frac{dQ_4}{ds} &=   -b_m(t-s)\beta_{mh}(Q_{1}Q_{4}-Q_{4}) -  \mu_m (t-s)(1-Q_4)-p \eta_b(t-s)\bigg{(} 1- \frac{L_{S}^{*}(t-s)}{K}  \bigg{)}Q_4(Q_5 - 1)\\
\notag\tiny \frac{dQ_5}{ds} &= -\eta_{l}(t-s)(Q_4-Q_5)-\mu_{l}(t-s)(1-Q_5),
\end{align}where $Q_k(0)=0,\ \forall k =1,\dots,5.$ 
Solving this system for every $t>0$ will enable us to compute $\tau \mapsto P_k(\tau, t), \ \forall \tau \in [0,t).$

\subsection{CTMC Model with periodical external introductions}
In this section, we will be considering external periodical introductions of infectious humans. These introductions aim to model infectious dynamics introduced in a completely susceptible local population by external infectious individuals, in a periodical environment.

For this, we consider $f: \mathbb{R_+} \rightarrow \mathbb{R_+}$ to be a nonnegative continuous $w-$periodical function. Thus, for this section, let us define the CTMC random variable as:
\[
X(t) = (S_h(t), E_h(t), I_h(t), I_{ext}(t), R_h(t), S_m(t), E_m(t), I_m(t), L_I(t), L_S(t)),
\] where $\forall t \geq 0,$ $X(t) \in \mathbb{Z}_{+}^{10}$. Therefore, following the previous subsection \ref{3.1}, this will lead us to the transition probabilities given by the following Table:

\begin{table}[h]
    \centering
    \begin{tabular}{|p{5.5cm}|c|p{6.5cm}|}
        \hline
        Description & Transition & Probability Rate \\
        \hline
        Introduction of external infectious humans & \( I_{ext} \to I_{ext}+1\) & \( f(t) \) \\
        Removal of external infectious humans & \( I_{ext} \to I_{ext}-1\) & \(\mu I_{ext} \) \\
        Recruitment of susceptible humans & \( S_h \to S_h + 1 \) & \( \mu_h N_h \) \\
        Death of susceptible humans & \( S_h \to S_h - 1 \) & \( \mu_h S_h \) \\
        Human infection via mosquito bite & \( S_h \to S_h - 1, E_h \to E_h + 1 \) & \( b_m(t) \beta_{mh} \frac{I_m S_h}{N_h} \) \\
        Progression from exposed to infectious humans & \( E_h \to E_h - 1, I \to I + 1 \) & \(  v_h E_h \) \\
        
        Recovery of infectious humans & \( I_h \to I_h - 1, R_h \to R_h + 1 \) & \( \eta_h I \) \\
        Death of exposed humans & \( E_h \to E_h - 1 \) & \( \mu_h E_h \) \\
        Death of infectious humans & \( I_h \to I_h - 1 \) & \( \mu_h I_h \) \\
        Death of recovered humans & \( R_h \to R_h - 1 \) & \( \mu_h R_h \) \\
        Recruitment of susceptible mosquitoes & \( S_m \to S_m + 1 \) & \( \eta_L(t) L_S \) \\
        Death of susceptible mosquitoes & \( S_m \to S_m - 1 \) & \( \mu_m(t) S_m \) \\
        Mosquito infection via biting humans & \( S_m \to S_m - 1, E_m \to E_m + 1 \) & \( b_m(t) \beta_{hm} \frac{(I_h +I_{ext}) S_m}{N_h} \) \\
        Progression from exposed to infectious mosquitoes & \( E_m \to E_m - 1, I_m \to I_m + 1 \) & \( \eta_m E_m \) \\
        Death of infectious mosquitoes & \( I_m \to I_m - 1 \) & \( \mu_m(t) I_m \) \\
        Recruitment of infected larvae & \( L_I \to L_I + 1 \) & \( \eta_b(t) \left(1 - \frac{L_I + L_S}{K} \right) p I_m \) \\
        Recruitment of susceptible larvae & \( L_S \to L_S + 1 \) & \( \eta_b(t) \left(1 - \frac{L_S + L_I}{K} \right) (S_m + E_m + (1 - p) I_m) \) \\
        Death of infected larvae & \( L_I \to L_I - 1 \) & \( \mu_L(t) L_I \) \\
        Death of susceptible larvae & \( L_S \to L_S - 1 \) & \( \mu_L(t) L_S \) \\
        \hline
    \end{tabular}
    \caption{Transition probabilities for the CTMC with periodical external introductions of infectious humans}
    \label{tab:transition}
\end{table}

\begin{table}[h]
    \centering
    \begin{tabular}{|l|c|c|}
        \hline
        Description & Transition & Probability Rate \\
        \hline
        External introduction of infectious humans & \(I_{ext} \to I_{ext}+1\)& f(t)\\
        Removal of external infectious humans & \(I_{ext} \to I_{ext}-1\) &\(\mu I_{ext}\)\\
        Human infection via mosquito bite & \( E_h \to E_h + 1 \) & \( b_m(t) \beta_{mh} I_m \) \\
        Progression from exposed to infectious humans & \( E_h \to E_h - 1, I_h \to I_h + 1 \) & \(  v_h E_h \) \\
        Recovery of infectious humans & \( I_h \to I_h - 1 \) & \( \eta_h I_h \) \\
        Death of exposed humans & \( E_h \to E_h - 1 \) & \( \mu_h I_h \) \\
        Death of infectious humans & \( I_h \to I_h - 1 \) & \( \mu_h I_h \) \\
        Mosquito infection via biting humans & \( E_m \to E_m + 1 \) & \( b_m(t) \beta_{hm} \frac{I S_m^{*}(t)}{N_h} \) \\
        Progression from exposed to infectious mosquitoes & \( E_m \to E_m - 1, I_m \to I_m + 1 \) & \( \eta_m E_m \) \\
        Death of infectious mosquitoes & \( I_m \to I_m - 1 \) & \( \mu_m(t) I_m \) \\
        Recruitment of infected larvae & \( L_I \to L_I + 1 \) & \( \eta_b(t) \left(1 - \frac{L_S^*(t)}{K} \right) p I_m \) \\ 
        Death of infected larvae & \( L_I \to L_I - 1 \) & \( \mu_L(t) L_I \) \\
        \hline
    \end{tabular}
    \caption{Transition probabilities for the Branching Process Approximation}
    \label{tab:BPA_ext}
\end{table}

Similarly to Section \ref{3.1}, we apply the Branching Process Approximation by assuming assumptions \ref{BPA_assumptions} and linearizing all transition rates, except for the introduction of external infectious humans rate, over the DFE (\ref{DFE}). Consequently, similar to \ref{3.1}, let $t >0$ and $\tau\in [0,t)$, and as the process is time-nonhomogeneous we define the time dependent transition probability from state $(i_1,i_2,i_3,i_4,i_5,i_6)\in \mathbb{Z}^{6}_{+}$ to $(j_1,j_2,j_3,j_4,j_5,j_6)\in \mathbb{Z}^{6}_{+}$, which keep track of the initial infectious variables introduction time $\tau>0$ and a final fixed time $t>\tau$, as \begin{align*} 
\mathbb{P}_{(i_1,i_2,i_3,i_4,i_5,i_6), (j_1,j_2,j_3,j_4,j_5,j_6)}(\tau, t) &= \mathbb{P}((E_{h}(t), I_h(t), I_{ext}(t), E_m (t), I_{m}(t), L_{I}(t))\\&=(i_1,i_2,i_3,i_4,i_5,i_6)|(E_{h}(t), I_h(t),I_{ext}(t), E_m (t), I_{m}(t), L_{I}(t))=(j_1,j_2,j_3,j_4,j_5,j_6)).
\end{align*}

$P^{i}(\tau, t):=\mathbb{P}_{(i_1,i_2,i_3,i_4,i_5,i_6), (0,\dots, 0)}(\tau, t) $ we obtain
\begin{align*}
-\frac{d}{d\tau} P^i(\tau, t)  =  &b_{m}(\tau )\beta_{mh}i_4(\mathbb{P}_{(i_1 +1  ,i_2,i_3,i_4,i_5,i_6), (0,\dots, 0)}(\tau, t)-P^{i}(\tau, t))  \\& +  \mu_h i_1(\mathbb{P}_{(i_1 -1 ,i_2,i_3,i_4,i_5,i_6), (0,\dots, 0)}(\tau, t)-P^{i}(\tau, t))  \\& + v_h i_1 (\mathbb{P}_{(i_1 -1,i_2 +1,i_3,i_4,i_5,i_6), (0,\dots, 0)}(\tau, t)-P^{i}(\tau, t))  \\& + (\eta_h + \mu_h) i_2 (\mathbb{P}_{(i_1 ,i_2 -1,i_3,i_4,i_5,i_6), (0,\dots, 0)}(\tau, t)-P^{i}(\tau, t)) \\&+ b_{m}(\tau  ) \beta_{hm}(i_2+i_3) \frac{S_{m}^{*}(\tau)}{N_h}(\mathbb{P}_{(i_1 ,i_2,i_3,i_4+1,i_5,i_6), (0,\dots, 0)}(\tau, t)-P^{i}(\tau, t)) \\& + f(t)  (\mathbb{P}_{(i_1 ,i_2,i_3 +1,i_4,i_5,i_6), (0,\dots, 0)}(\tau, t)-P^{i}(\tau, t))
\\&\ +\mu \ i_3 (\mathbb{P}_{(i_1 ,i_2,i_3 -1,i_4,i_5,i_6), (0,\dots, 0)}(\tau, t)-P^{i}(\tau, t)) 
\\&\ + \mu_{m}(\tau) i_4 (\mathbb{P}_{(i_1 ,i_2,i_3,i_4 -1,i_5,i_6), (0,\dots, 0)}(\tau, t)-P^{i}(\tau, t)) 
\\&\  + \eta_m(\tau) i_5 (\mathbb{P}_{(i_1,i_2,i_3,i_4-1,i_5+1,i_6), (0,\dots, 0)}(\tau, t)-P^{i}(\tau, t)) 
\\&\ +\mu_m (\tau ) i_6 (\mathbb{P}_{(i_1,i_2,i_3,i_4,i_5 - 1,i_6), (0,\dots, 0)}(\tau, t) -P^{i}(\tau, t))
\\&\ +p\ i_5  \eta_b (\tau)\big( 1 - \frac{L_{S}^{*}(\tau) }{K}  \big)(\mathbb{P}_{(i_1,i_2,i_3,i_4,i_5,i_6+1), (0,\dots, 0)}(\tau, t)-P^{i}(\tau, t))
\\&\ + \eta_{l}(\tau )i_6 (\mathbb{P}_{(i_1,i_2,i_3,i_4,i_5 +1,i_6 -1), (0,\dots, 0)}(\tau, t) -P^{i}(\tau, t))
\\&\ +\mu_{l}(\tau)i_6 (\mathbb{P}_{(i_1,i_2,i_3,i_4,i_5,i_6 -1), (0,\dots, 0)}(\tau, t)-P^i(\tau, t)).
\end{align*}

Define \begin{align*}  P_1(\tau, t) &=  \mathbb{P}_{(1,0,0,0,0,0), (0,\dots, 0)}(\tau, t), \\P_2(\tau, t)  &=  \mathbb{P}_{(0,1,0,0,0,0), (0,\dots, 0)}(\tau, t), \\ P_3(\tau, t) &=  \mathbb{P}_{(0,0,1,0,0,0), (0,\dots, 0)}(\tau, t),\\ P_4(\tau, t) &=  \mathbb{P}_{(0,0,0,1,0,0), (0,\dots, 0)}(\tau, t),\\ P_5(\tau, t) &=  \mathbb{P}_{(0,0,0,0,1,0), (0,\dots, 0)}(\tau, t),\\ P_6(\tau, t) &=  \mathbb{P}_{(0,0,0,0,0,1), (0,\dots, 0)}(\tau, t).
\end{align*}%Moreover, as $f(t)$ has compact support, then we have that for $t>0$ sufficiently large, and any $\tau <t,$ $P_3(\tau, t)=0.$

Now, due to the Branching Process Approximation Assumptions (i)-(iii), we assume that \begin{align*}
    \mathbb{P}_{(1,0,0,0,1,0), (0,\dots, 0)}(\tau, t)&= P_{1}(\tau, t)P_5(\tau, t),\\
    \mathbb{P}_{(0,0,0,0,1,1), (0,\dots, 0)}(\tau, t)&= P_{6}(\tau, t)P_5(\tau, t),\\
    \mathbb{P}_{(0,1,0,1,0,0), (0,\dots, 0)}(\tau, t)&= P_{2}(\tau, t)P_4(\tau, t),\\
    \mathbb{P}_{(0,1,1,0,0,0), (0,\dots, 0)}(\tau, t)&= P_{2}(\tau, t)P_3(\tau, t),\\ \mathbb{P}_{(0,0,1,1,0,0), (0,\dots, 0)}(\tau, t)&= P_{3}(\tau, t)P_4(\tau, t),\\ \mathbb{P}_{(1,0,1,0,0,0), (0,\dots, 0)}(\tau, t)&= P_{3}(\tau, t)P_1(\tau, t),\\ \mathbb{P}_{(0,0,1,0,1,0), (0,\dots, 0)}(\tau, t)&= P_{3}(\tau, t)P_5(\tau, t),\\ \mathbb{P}_{(0,0,1,0,0,1), (0,\dots, 0)}(\tau, t)&= P_{3}(\tau, t)P_6(\tau, t),\\\mathbb{P}_{(0,0,2,0,0,0), (0,\dots, 0)}(\tau, t)&= P_{3}^2(\tau, t).
\end{align*} 

These assumptions will allow us to obtain the 6-dimensional system: 

\begin{align*}
\tiny
\frac{dP_1}{d\tau}(t,\tau) &= -\mu_h(1-P_1(\tau,t)) -v_h (P_2 (\tau, t) - P_1(\tau, t)) - f(\tau)(P_1(\tau, t)P_3(\tau, t)-P_1(\tau, t))\\
\tiny \frac{dP_2}{d\tau}(t,\tau) &= -(\eta_h + \mu_h)(1-P_2(\tau, t)) - b_{m}(\tau)\beta_{hm}\frac{S_{m}^{*}(\tau)}{N_h}(P_3(\tau, t)P_{2}(\tau, t)-P_{2}(\tau, t)) - f(\tau)(P_2(\tau, t)P_3(\tau, t)-P_2(\tau, t))\\
\frac{dP_3}{d\tau}(t,\tau) &= -f(\tau)(P_3^2(\tau,t)-P_3(\tau, t))-\mu (1-P_3(\tau,t))
\\
\tiny \frac{dP_4}{d\tau}(t,\tau) &= -\mu_m(\tau)(1-P_4(\tau, t)) - \eta_m(\tau)(P_5(\tau, t)-P_4(\tau,t))- f(\tau)(P_4(\tau, t)P_3(\tau, t)-P_4(\tau, t))\\
\tiny\frac{dP_5}{d\tau}(t,\tau) &=   -b_m(\tau)\beta_{mh}(P_{1}(\tau, t)P_{5}(\tau, t)-P_{5}(\tau,t)) -  \mu_m (\tau)(1-P_5(\tau, t))\\&-p \eta_b(\tau)\bigg{(} 1- \frac{L_{S}^{*}(\tau)}{K}  \bigg{)}P_5(\tau, t)(P_6(\tau, t) - 1)- f(\tau)(P_5(\tau, t)P_3(\tau, t)-P_5(\tau, t))\\
\frac{dP_6}{d\tau}(t,\tau) &= -\eta_{l}(\tau)(P_5(\tau, t)-P_6(\tau, t))-\mu_{l}(\tau)(1-P_6(\tau, t))- f(\tau)(P_6(\tau, t)P_3(\tau, t)-P_6(\tau, t)),\end{align*}for $t>0$ fixed and $\tau \in [0,t)$.

Now let $s=t - \tau$ and define $Q_k (s) = P_k (t-s,t),$ then we arrive in the following system
\begin{align}\label{ext_prob_intro}
\frac{dQ_1}{ds} &= -\mu_h(1-Q_1) -v_h (Q_2  - Q_1)- f(t-s)(Q_1 Q_3-Q_1)\\
\notag\tiny \frac{dQ_2}{ds} &= -(\eta_h + \mu_h)(1-Q_2) - b_{m}(t-s)\beta_{hm}\frac{S_{m}^{*}(t-s)}{N_h}(Q_3 Q_{2}-Q_{2})- f(t-s)(Q_2 Q_3-Q_2)\\
\notag \frac{dQ_3}{ds} &= -f(t-s)(Q_3^2-Q_3)-\mu (1-Q_3)
\\
\notag\tiny \frac{dQ_4}{ds} &= -\mu_m(t-s)(1-Q_4) - \eta_m(t-s)(Q_5-Q_4)- f(t-s)(Q_4 Q_3-Q_4)\\
\notag\tiny \frac{dQ_5}{ds} &=   -b_m(t-s)\beta_{mh}(Q_{1}Q_{5}-Q_{5}) -  \mu_m (t-s)(1-Q_5)\\\notag &-p \eta_b(t-s)\bigg{(} 1- \frac{L_{S}^{*}(t-s)}{K}  \bigg{)}Q_5(Q_6 - 1)- f(t-s)(Q_5 Q_3-Q_5)\\
\notag\tiny \frac{dQ_6}{ds} &= -\eta_{l}(t-s)(Q_5-Q_6)-\mu_{l}(t-s)(1-Q_6)- f(t-s)(Q_6 Q_3-Q_6),
\end{align}where $Q_k(0)=0,\ \forall k =1,\dots,6.$ 
Solving this system for every $t>0$ will enable us to compute $\tau \mapsto P_k(\tau, t), \ \forall \tau \in [0,t).$

\section{Numerical Analysis}
\label{sec:numerical}
\subsection{Model parameters}
In this section we simulate the CTMC process and plot trajectories for different initial conditions, taking into account different biting rate levels. The following table lists the temperature dependent parameters and constants used in the present model.

\begin{table}[htbp]\label{table}
\centering
\caption{Features and Parameters in the Model}
\label{table}
\begin{tabular}{|c|c|}
\hline
\textbf{Parameter} & \textbf{Constants and Temperature Dependent Formula} \\
\hline
$\mu_{h}$ & $\frac{1}{74.5*365} $ ${\rm day}^{-1}$ \cite{WorldBank2024}\\ 
$N_{h}$ & $2700000$ \cite{WorldPopulationReview2024}
\\

$\mu_{m}(T)$ & $0.8692 - 0.1590T + 0.01116(T^{2}) -0.0003408(T^{3})+3.809(10^{-6})T^{4}\ {\rm day}^{-1} $ \cite{McLean1974}
\\
$b_{m}(T)$ & $0.0943 + 0.0043 * T \ {\rm day}^{-1}$ \cite{Focks1995} \\

$\beta_{mh}$ & $0.13$ \cite{Liu2020} \\
$\beta_{hm}$ & $0.3$ \cite{Liu2020} \\ 
$\mu_{l}(T)$ & $2.130 - 0.3797 \cdot T + 0.02457 \cdot T^2 - 0.0006778 \cdot T^3 + 0.000006794 \cdot T^4\ {\rm day}^{-1}$ \cite{Valdez2018} \\ 
$\eta_h$ & $5\ day$ \cite{CDCChikungunyaSymptoms} \\ 
$v_{h}$ & $0.2 \  {\rm day^{-1}},$ \cite{CDC} \\\\
 $\eta_l(T)$ & $\frac{0.00070067 \cdot (T + 273.15) \cdot e^{13093 \cdot \left(\frac{1}{298} - \frac{1}{T + 273.15}\right)}}{1 + e^{28715 \cdot \left(\frac{1}{304.6} - \frac{1}{T + 273.15}\right)}}$ \cite{Liu2020}\\\\
$\eta_{b}(T)$ & $-0.32 + 0.11 \cdot T - 0.12 \cdot T^{2} + 0.01 \cdot T^{3} - 0.00015 \cdot T^{4} \ {\rm day}^{-1} $ \cite{Valdez2018} \\
\hline
\end{tabular}
\end{table}
Given the temperature data collected from \cite{power_dav}, it was utilized Fast Fourier Transform from SciPy.FFT library in Python \cite{gondim2025} to estimate the periodic and time dependent average temperature functions in Miami, $T_{MIA}(t)$, defined as:

\[
T_{MIA}(t) =
3.6375\cdot\sin\left(\frac{2\pi (t+59)}{365} + 3.7\right)+25.91,
\] %somo 59 dias pq foi quando a covid comecou, esse eh o ida em q inicializo o modelo!!!!!
where the unit of the time $t$ is $day$, and the temperature unit is Celsius ($^{\circ}C$).
The use of air conditioning not only makes the environment less accessible to mosquitoes but also affects their survival and distribution, leading to lower biting rates and reduced disease transmission in the U.S. compared to other countries, as addressed in \cite{reiter2003texas}. Thus, we consider the per-capita biting rate in Miami as: $$
b_m^{MIA}(t) = 0.07544 + 0.00344 \cdot T_{MIA}(t).
$$In the next two subsections \ref{4.2} and \ref{4.3}, we will perform a sensitivity analysis of different parameters and understand how the probability of disease extinction will behave as we vary these parameters. Lastly, we will be plotting the CTMC trajectories and comparing them to the disease extinction probability.

\subsection{Numerical Analysis without periodical external introductions  }\label{4.2}

Consider the positive constants $b_{mult}, \in (0,1.25]$ and $\alpha, \in (0,1.2]$. These will be referred as the biting rate and carrying capacity multipliers, respectively. In the simulations we shall be considering $\alpha  K$ and $b_{m_{mult}}  b_{m}(t)$ for different values of the multipliers, aiming to vary these parameters so to understand how different control strategies and distinct scenarios can influence the model outcomes. 
  
Let us fix a final time for the disease extinction probability to be 2555 days, which approximately 7 years, from now on. The selection of such number of years for our numerical analysis is motivated by the need to observe the long-term behavior of the stochastic system under study. While the specific duration of 5, 7, or 10 years is somewhat arbitrary, a multi-year period is essential to capture the full effect of numerous seasonal cycles on transmission dynamics and to provide a sufficiently long window for the probability of disease extinction to approach its asymptotic value.

\begin{figure}[htbp] 
    \centering
    \begin{subfigure}    
    [b]{0.45\textwidth}
        \centering
        \includegraphics[width=\textwidth]{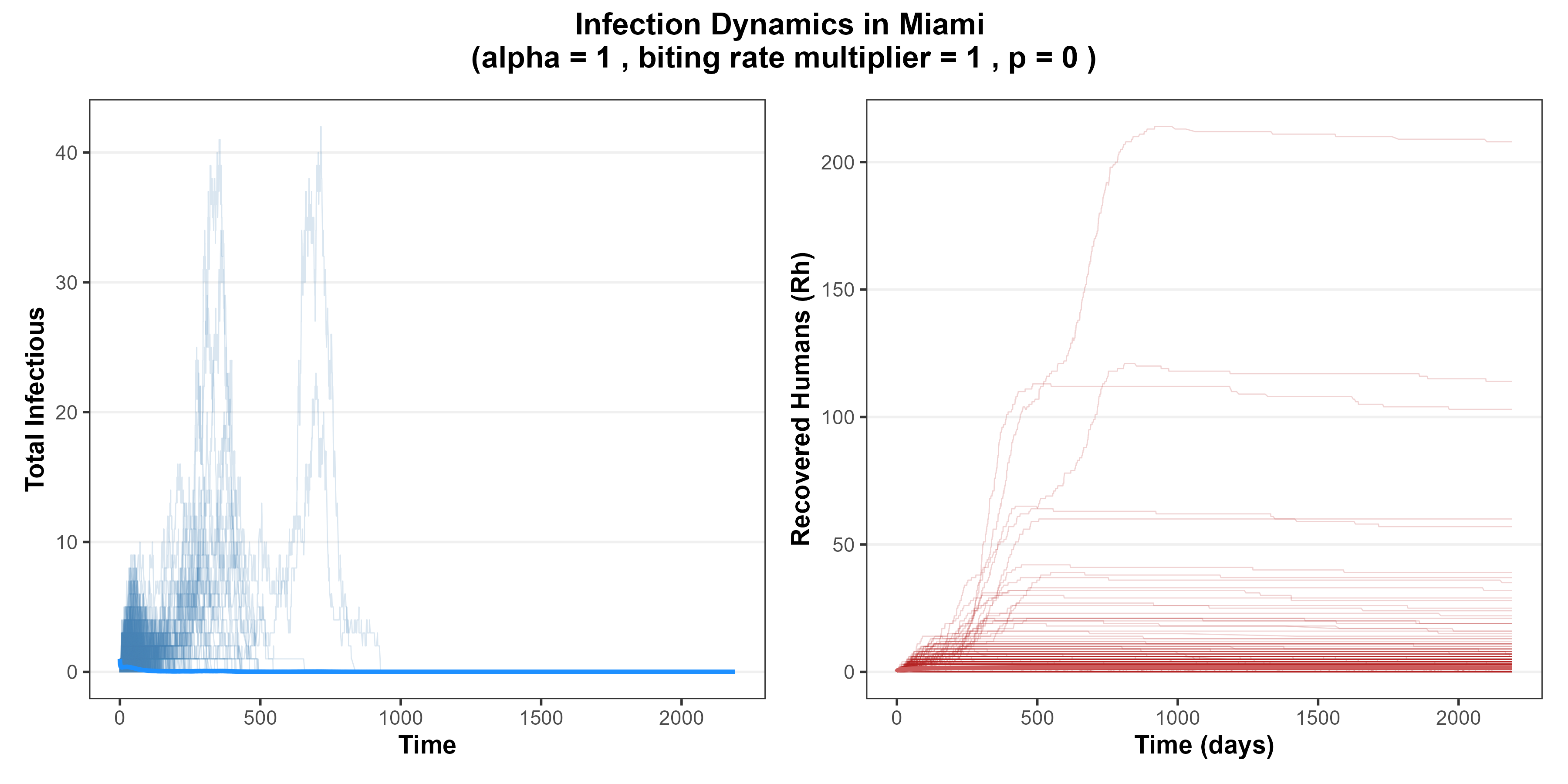}
        \caption{ p=0}
        \label{fig:plot1}
    \end{subfigure}
    \hfill
    \begin{subfigure}[b]{0.45\textwidth}
        \centering
        \includegraphics[width=\textwidth]{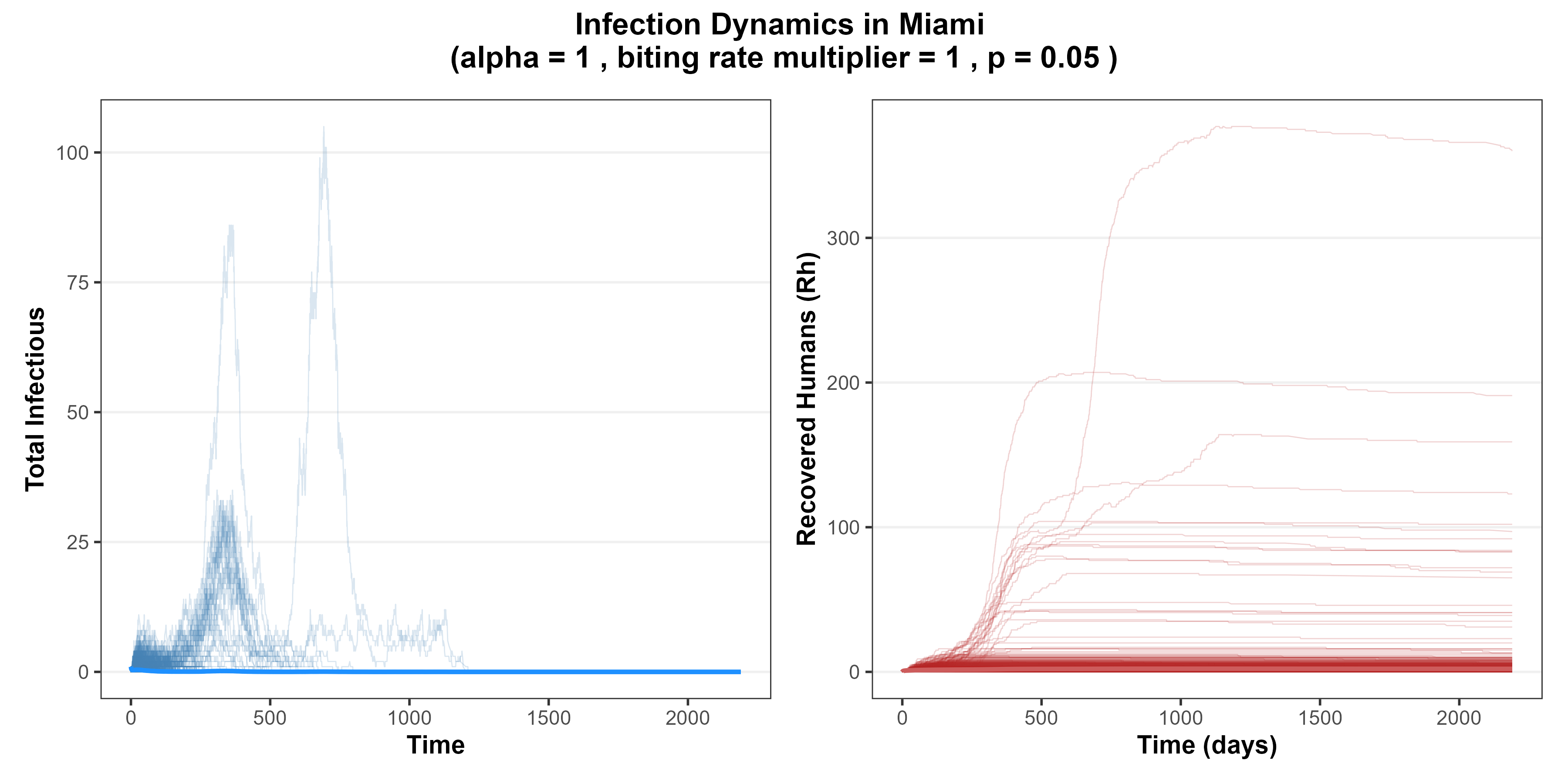}
        \caption{ p=0.05}
        \label{fig:plot2}
    \end{subfigure}
    \vspace{0.05cm}
    \begin{subfigure}[b]{0.45\textwidth}
        \centering
        \includegraphics[width=\textwidth]{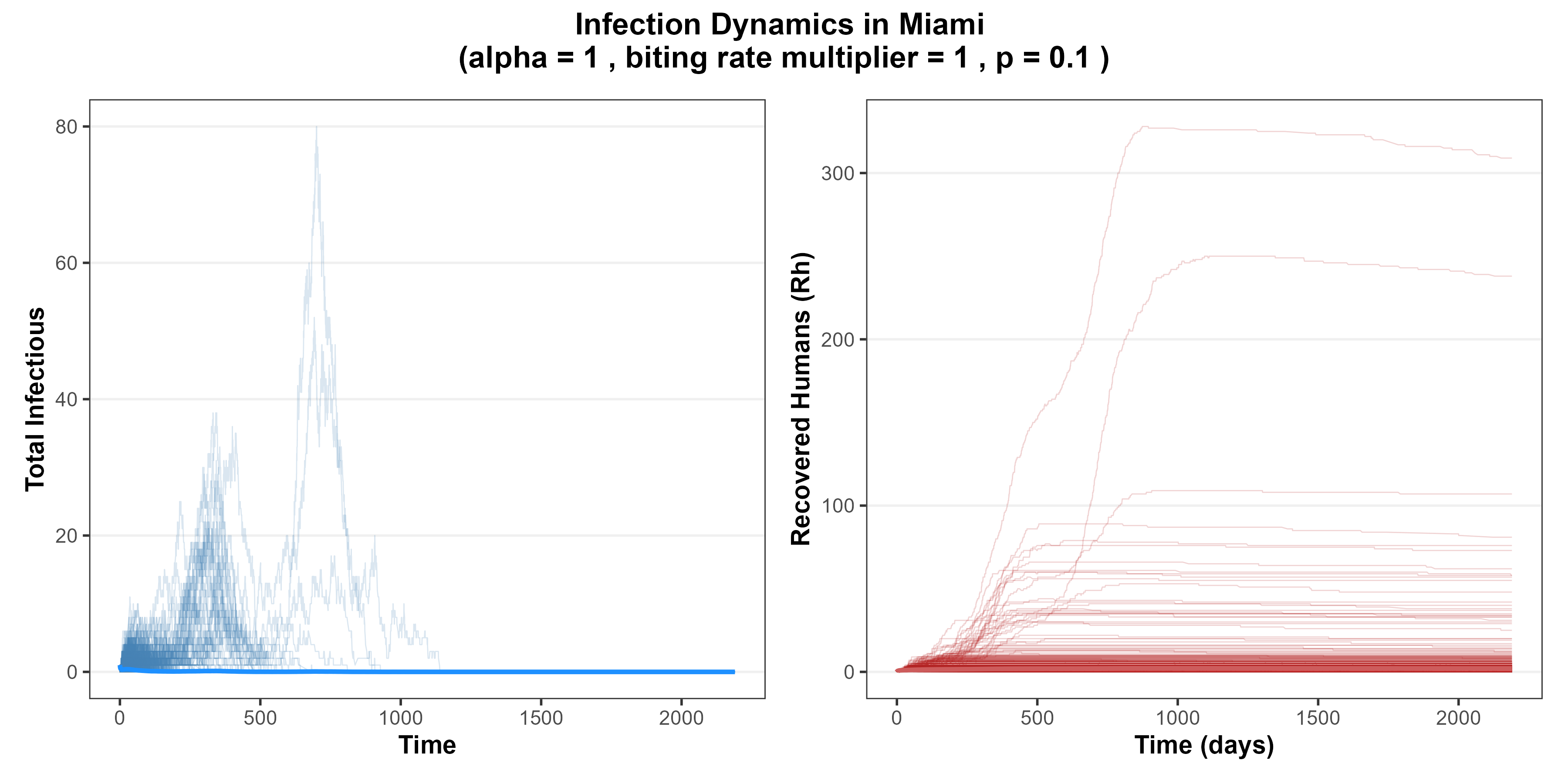}
        \caption{p=0.1 }
        \label{fig:plot3}
    \end{subfigure}
    \hfill
    \begin{subfigure}[b]{0.45\textwidth}
        \centering
        \includegraphics[width=\textwidth]{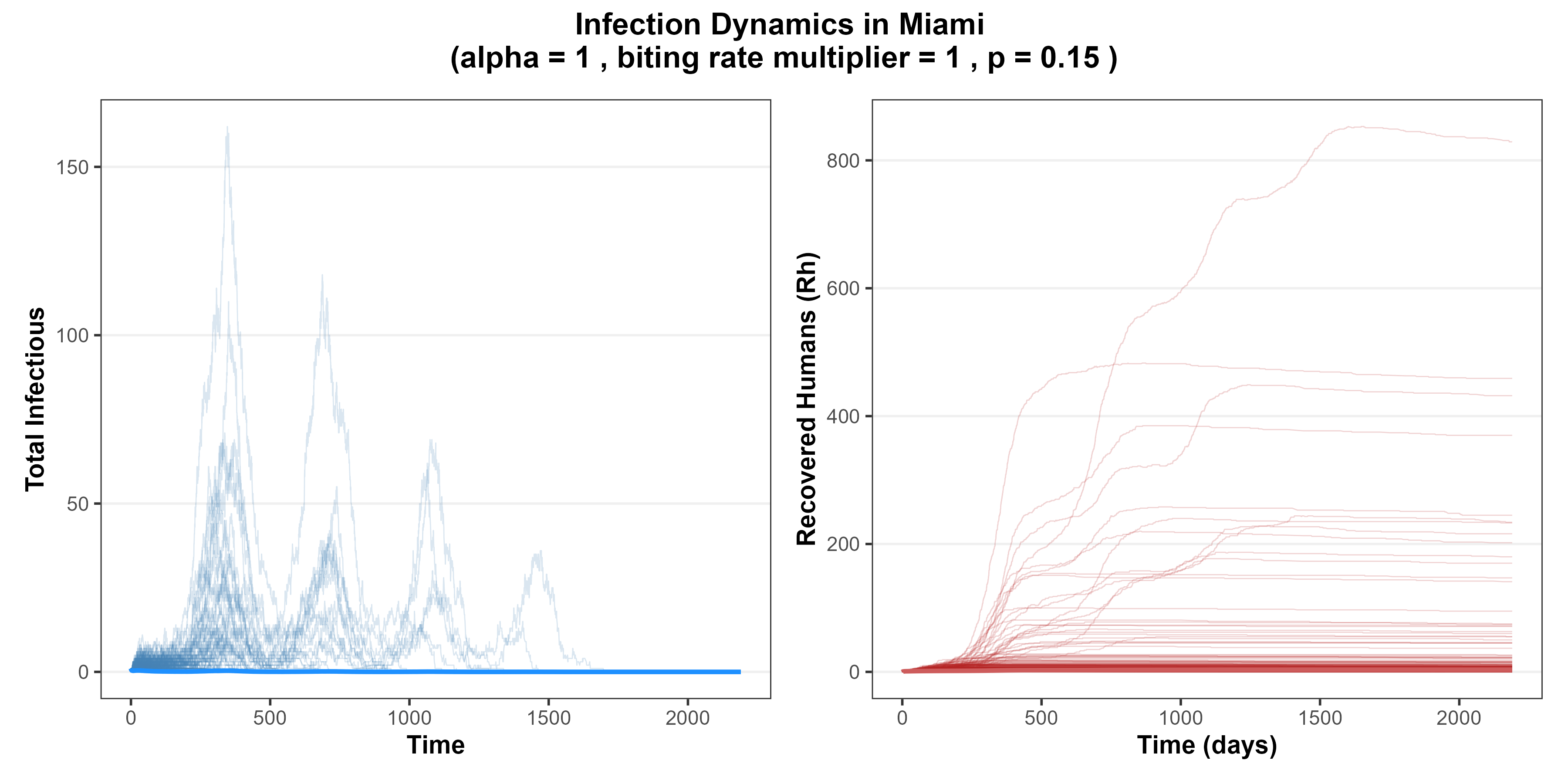}
       \caption{p=0.15 }
        \label{fig:plot4}
    \end{subfigure} 
    \caption{Infectious dynamics for different values of vertical probability $p \in [0,0.15]$, given an introduction of one infectious individual on January $1^{st}$, which is the initial time on the plot. In total, 2000 trajectories are plotted and for each figure the solid lines represent the mean.}
    \label{fig:five_plots0}
\end{figure}

First, let us simulate the CTMC whose rates are defined in Table \ref{tab:transition}, and considering an initial introduction of one individual on January $1^{st}$, and varying $p \in [0, 0.15]$, we obtain Figure \ref{fig:five_plots0}.


In light of the assumptions from Section 3 and considering System (\ref{ext_prob}), let us define the probability of disease extinction within the first 2555 days as $P_{extinction}(\tau):=P_{(0,1,0,0,0),(0,\dots,0)}(\tau, 2555),\ \forall \tau \in [0,2555)$. Hence, essentially, there will be an introduction of one infectious individual at time $\tau>0,$ and the probability for the disease to die out is computed at the final time $t=2555,$ or after 7 years.
\begin{figure}[b]
    \centering
    \includegraphics[width=\textwidth]{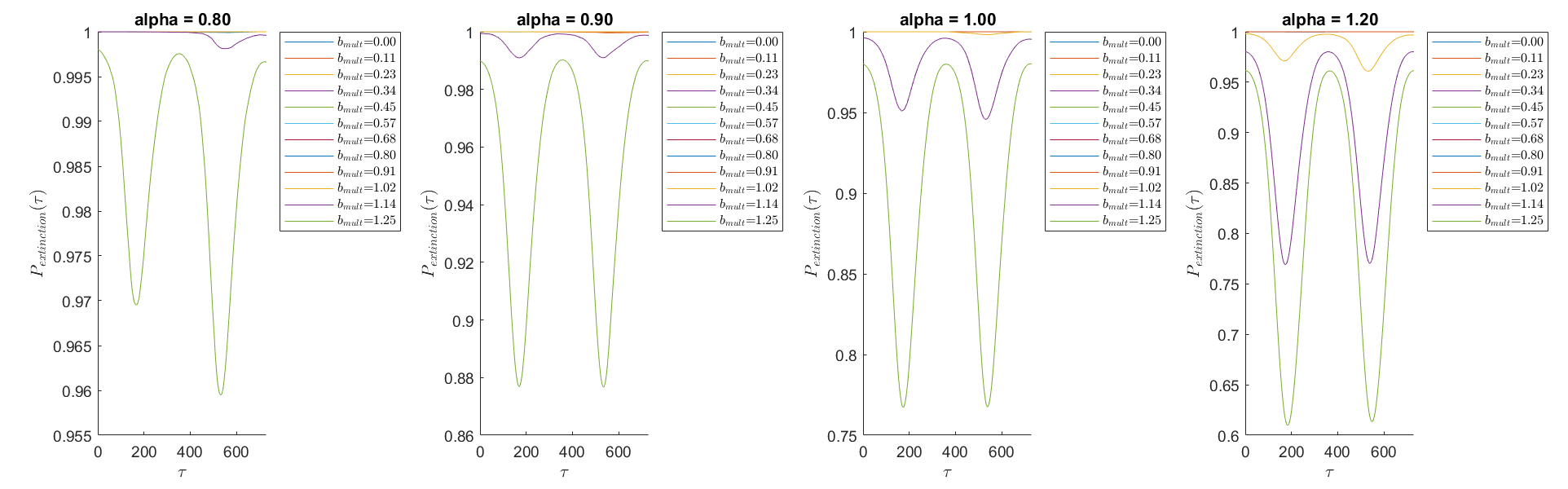}
    \caption{Probability of disease extinction given a single introduction of one external individual at each time $\tau \in [0, 2555)$ in Miami. A sensitivity analysis of the biting rate and carrying capacity multipliers was performed.}
    \label{prob_ALL_one_intro}
\end{figure}
\pagebreak

Figure \ref{prob_ALL_one_intro} shows the importance of reducing the carrying capacity and biting rate multipliers as the main ways to adopt control strategies to avoid a disease outbreak, given the external introduction of an infectious individual into the local community. It also shows the times of the year where the probability of disease extinction within 7 years is lower, i.e. the most propense times of the year where an external introduction of an infectious individual could potentially lead to an outbreak- across all of the probability curves that are not approaching the constant function equal to one-are in between June $27^{th}$ and July $7^{th}$. Moreover, similarly, the highest probability extinction values within the first 2555 days occur when an external introduction of an infectious individual is made at the beginning of the year, around January $1^{st}.$

\begin{figure}[htbp]
    \centering
    \begin{subfigure}[c]{0.45\textwidth}
        \centering
        \includegraphics[width=\textwidth]{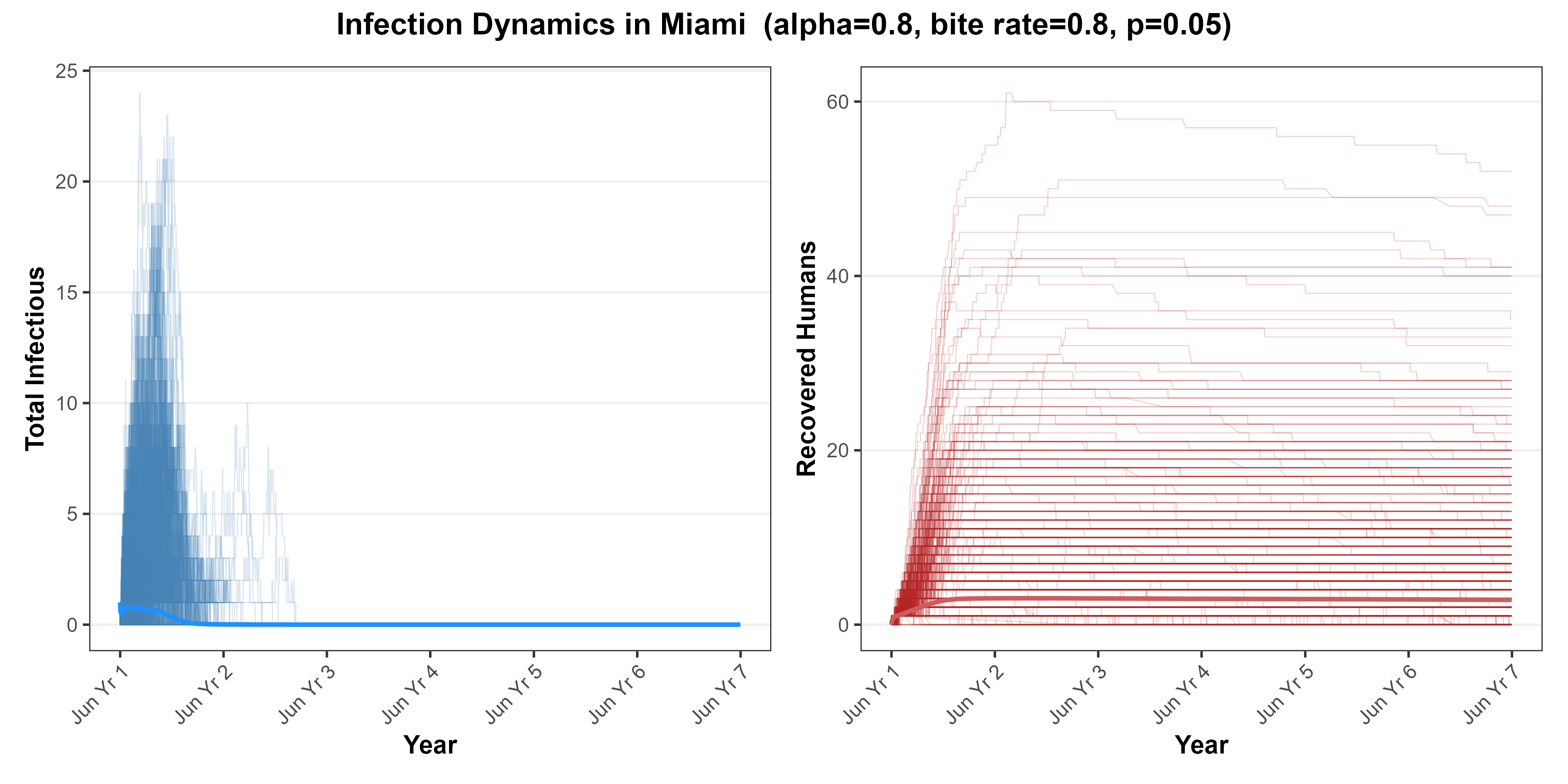}
        \caption{ }
        \label{jun_sens_analy_one_introA}
    \end{subfigure}
    \hfill
    \begin{subfigure}[c]{0.45\textwidth}
        \centering
        \includegraphics[width=\textwidth]{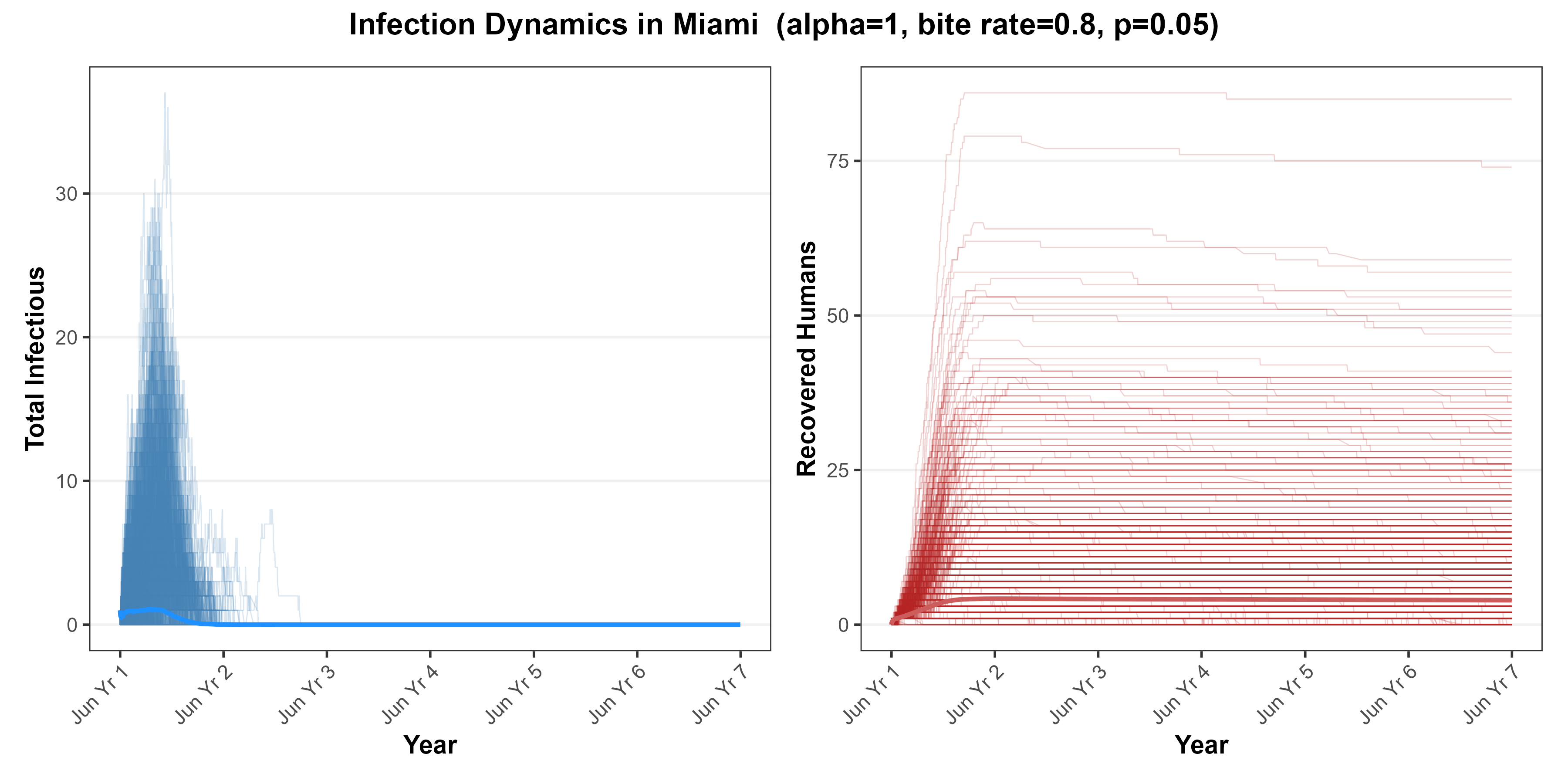}
        \caption{}
        \label{}
    \end{subfigure}
    \vspace{0.05cm}
    \begin{subfigure}[b]{0.45\textwidth}
        \centering
        \includegraphics[width=\textwidth]{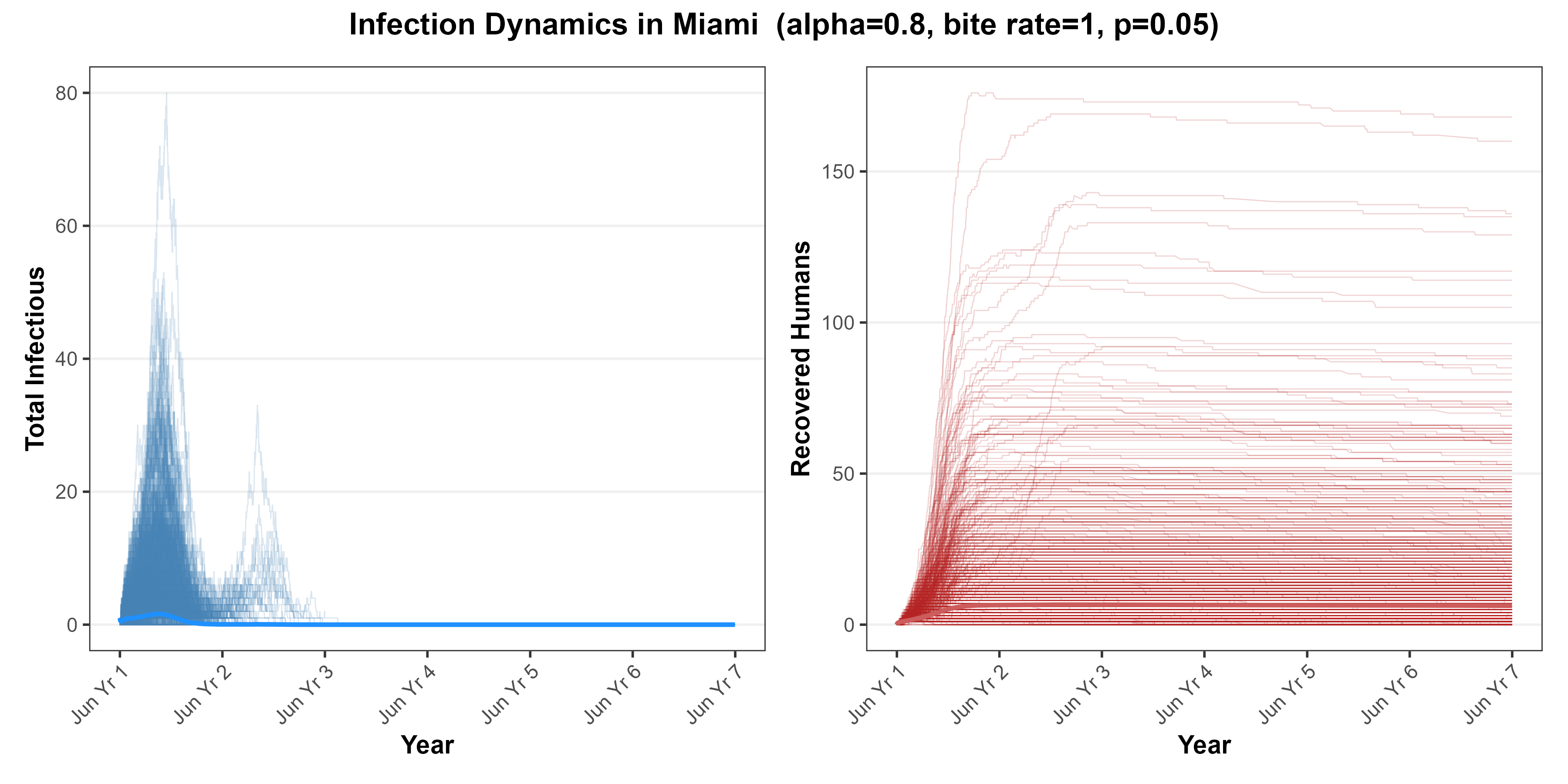}
        \caption{}
        \label{jun_sens_analy_one_introZ}
    \end{subfigure}
    \caption{Infectious dynamics for different values of the biting rate multiplier $b_{mult}$ and the carrying capacity multiplier $\alpha$, given an introduction of one infectious individual on June $29^{th}$ which is the initial time on the plot. In total, 2000 trajectories are plotted and for each figure the solid lines represent the mean.}
    \label{fig:june_five_plots}
\end{figure}

Now, considering the sensitivity analysis performed in Figure \ref{prob_ALL_one_intro}, and following \cite{gondim2025}, we fix $p =0.05.$ Let us understand how the disease dynamics-based on the CTMC Table \ref{tab:transition}- will change as we vary the multipliers of biting rate and carrying capacity, for an external introduction occurring on January $1^{st}$ and June $29^{th}$, individually; by analyzing the following Figures \ref{jan_sens_analy_one_introA} through \ref{jan_sens_analy_one_introZ} and Figures \ref{jun_sens_analy_one_introA} through \ref{jun_sens_analy_one_introZ}, respectively.

\begin{figure}[htbp]
    \centering
    \begin{subfigure}[c]{0.45\textwidth}
        \centering
        \includegraphics[width=\textwidth]{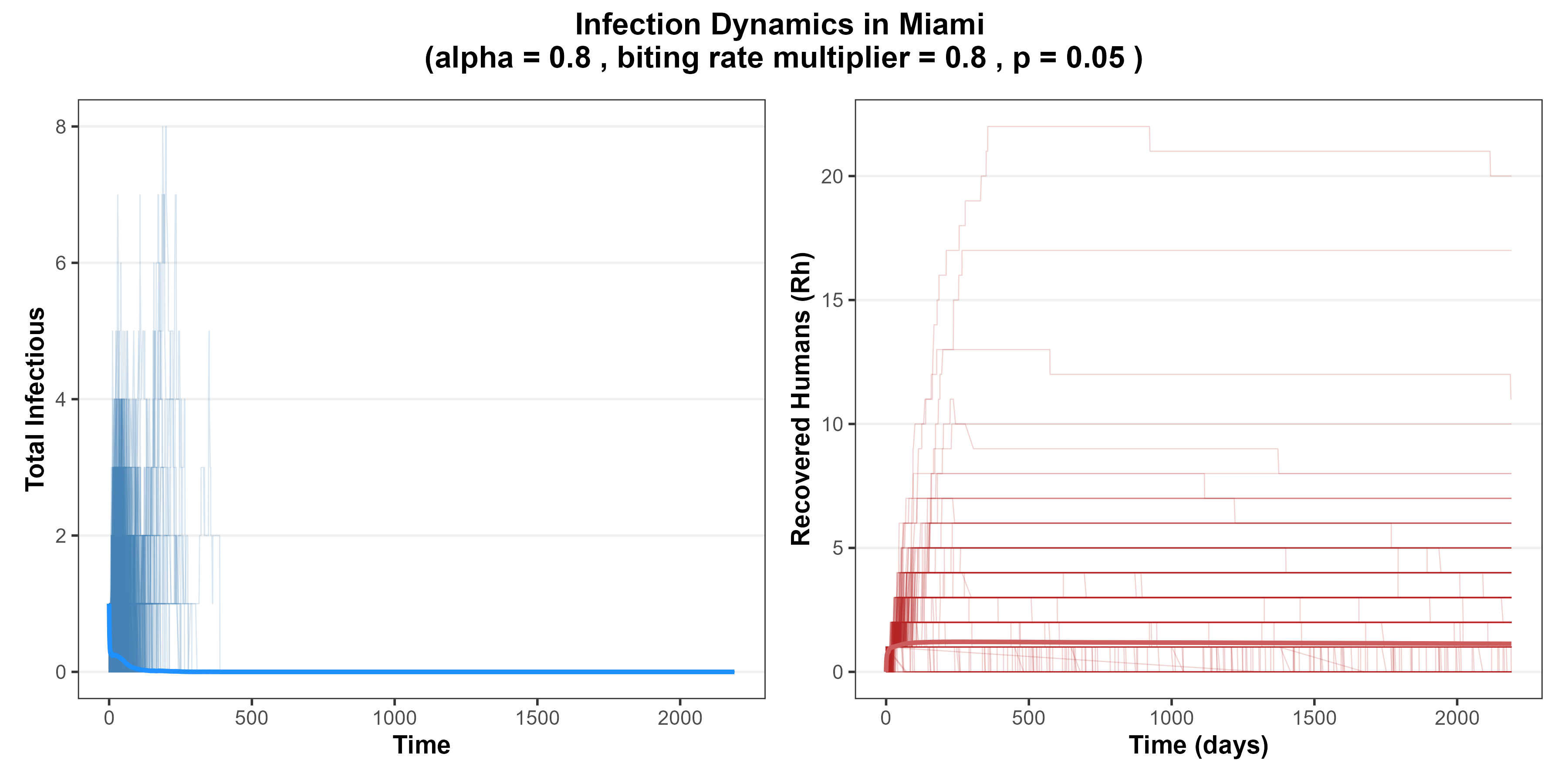}
        \caption{ }
        \label{jan_sens_analy_one_introA}
    \end{subfigure}
    \hfill
    \begin{subfigure}[c]{0.45\textwidth}
        \centering
        \includegraphics[width=\textwidth]{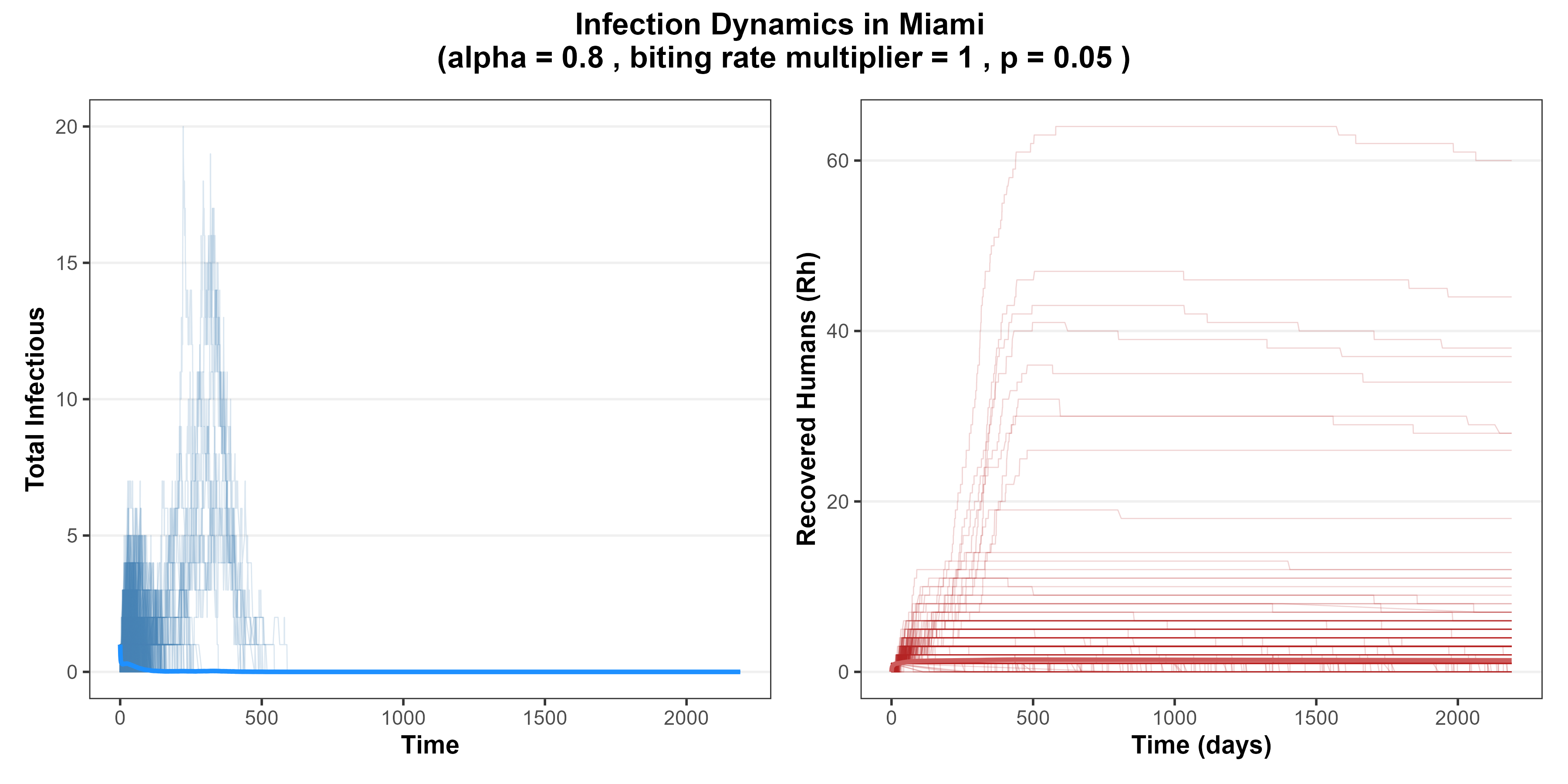}
        \caption{}
        \label{}
    \end{subfigure}
    \vspace{0.05cm}
    \begin{subfigure}[b]{0.45\textwidth}
        \centering
        \includegraphics[width=\textwidth]{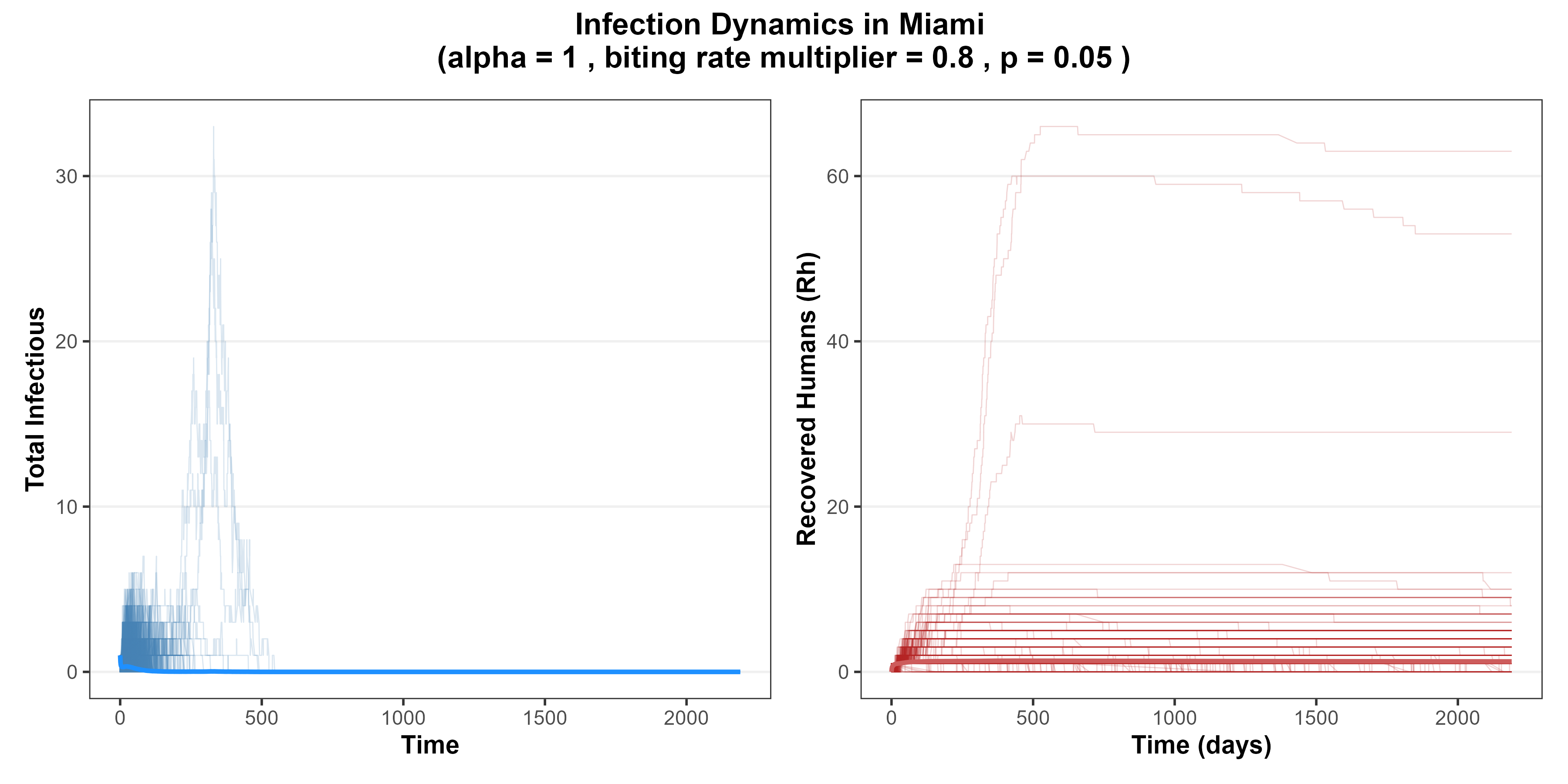}
        \caption{}
        \label{jan_sens_analy_one_introZ}
    \end{subfigure}
    \caption{Infectious dynamics for different values of the biting rate multiplier $b_{mult}$ and the carrying capacity multiplier $\alpha$, given an introduction of one infectious individual on January $1^{st}$ which is the initial time on the plot. In total, 2000 trajectories are plotted and for each figure the solid lines represent the mean.}
    \label{fig:jan_five_plots}
\end{figure}

\pagebreak

\begin{figure}
        \begin{subfigure}[c]{0.25\textwidth}
        \centering
        \includegraphics[width=\textwidth]{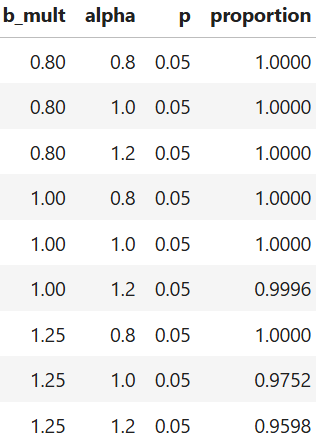}
        \caption{}
        \label{test_prob_jan}
    \end{subfigure}
    \hspace{1 cm}
    \begin{subfigure}[c]{0.25\textwidth}
        \centering
        \includegraphics[width=\textwidth]{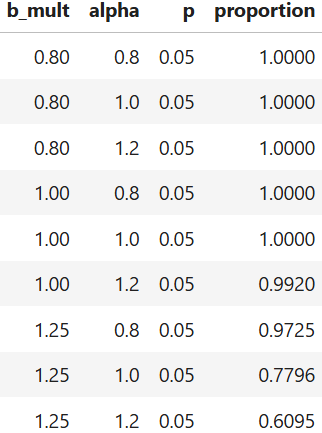}
        \caption{}
        \label{test_prob_june}
    \end{subfigure}
       \caption{Probability of disease extinction given a single introduction of one external individual at each time $\tau \in [0, 2555)$ in Miami-Dade can also be computed by calculating the proportion of trajectories that reached the zero state within 7 years, considering an initial single introduction on January $1^{st}$ (a) and June $29^{th}$ (b) and a total of 5000 trajectories (or samples) obtained by simulating the CTMC. The probability of vertical transmission is set to be a constant and the carrying capacity and biting rate multipliers are allowed to vary. }
\end{figure}

In the following, due to the simulations obtained from the CTMC defined in Section 3, whose transition probabilities are defined in Table \ref{tab:transition}, it was possible to construct the Table \ref{test_prob_jan} and Table \ref{test_prob_june} through using the CTMC to construct 5000 trajectories to show the proportion of time series that reached the zero state after 2555 days after the introduction of one infectious individual in the initial day. Tables \ref{test_prob_jan} and \ref{test_prob_june} will be compared against the probability of disease extinction within 2555 days, given initial introductions of an infectious individual at $t=0$ and $t=179$-i.e, initial introduction of an infectious individual on January $1^{st}$ and June $29^{th}$, respectively-, individually, computed by the Branching Process Approximation assumptions considered in Section 3 and whose probability of extinction is shown in Figure \ref{prob_ALL_one_intro}, for each value of vertical probability, biting rate and carrying capacity multiplier. Thus, Table \ref{test_prob_jan} and Table \ref{test_prob_june} can be considered as an hypothesis testing for the Branching Process Approximation assumptions assumed in Section 3. 

Indeed, we notice the proportion of trajectories reaching the zero state within 7 years established from the CTMC and the probability of disease extinction values, for each value of carrying capacity and biting rate multipliers, for the initial introduction days January $1^{st}$ and June $29^{th}$ are comparable, indicating the assumptions assumed in the Branching Process Approximation are reliable for the computation of the probability of disease extinction.

\pagebreak
\subsection{Numerical Analysis with periodical external introductions}\label{4.3}
In the subsection, let us focus on the computational analysis of multiple introductions of external infectious individuals in Miami-Dade.
Multiple introductions will be divided into sparsely distributed introductions and densely distributed introductions, which mean distinct introductions of a fixed number of infected individuals during a short amount of time-e.g, throughout a few days-, and during a larger amount of time-e.g, in distinct months-. The external infectious individuals introduction function is a piece-wise periodic constant function,  used in this analysis is defined by the following algorithm:\begin{algorithm}
\caption{Generalized External Introduction Function}
\begin{algorithmic}[1]
\REQUIRE $t$: time, $T$: period (default 365), $\tau_{\text{tol}}$: tolerance (default 0.5)
\REQUIRE $\mathbf{t}_{\text{intro}}$: vector of introduction times
\ENSURE $y$: output value (0 or 1)
\STATE $\tau \gets t \bmod T$
\STATE $y \gets 0$
\FOR{$i = 1$ to $\text{length}(\mathbf{t}_{\text{intro}})$}
    \IF{$|\tau - \mathbf{t}_{\text{intro}}[i]| < \tau_{\text{tol}}$}
        \STATE $y \gets 1$ \COMMENT{At most one infected individual inserted per day}
        \STATE \textbf{break}
    \ENDIF
\ENDFOR
\RETURN $y$
\end{algorithmic}
\end{algorithm}

Where the pattern of introductions is determined by $\mathbf{t}_{\text{intro}}$:
\begin{itemize}
    \item For sparsely distributed introductions: $\mathbf{t}_{\text{intro}}$ contains values with large intervals (e.g., $[11, 53, 94, 125]$)
    \item For densely distributed introductions: $\mathbf{t}_{\text{intro}}$ contains values with small intervals (e.g., $[2, 3, 4, 5]$)
\end{itemize}

The function returns 1 (indicating introduction of one infected individual) when the time modulo period $T$ is within tolerance $\tau_{\text{tol}}$ of any introduction time in $\mathbf{t}_{\text{intro}}$, and 0 otherwise. We consider $T=365$, i.e, its period is set to be one year (in days).

Now,considering the probability of disease extinction computed by solving System \ref{ext_prob_intro}, for densely and sparsely distributed introductions made from January to June, respectively, we obtain the following Figure \ref{dense_intro}.

\begin{figure}
        \begin{subfigure}[c]{0.45\textwidth}
        \centering
        \includegraphics[width=\textwidth]{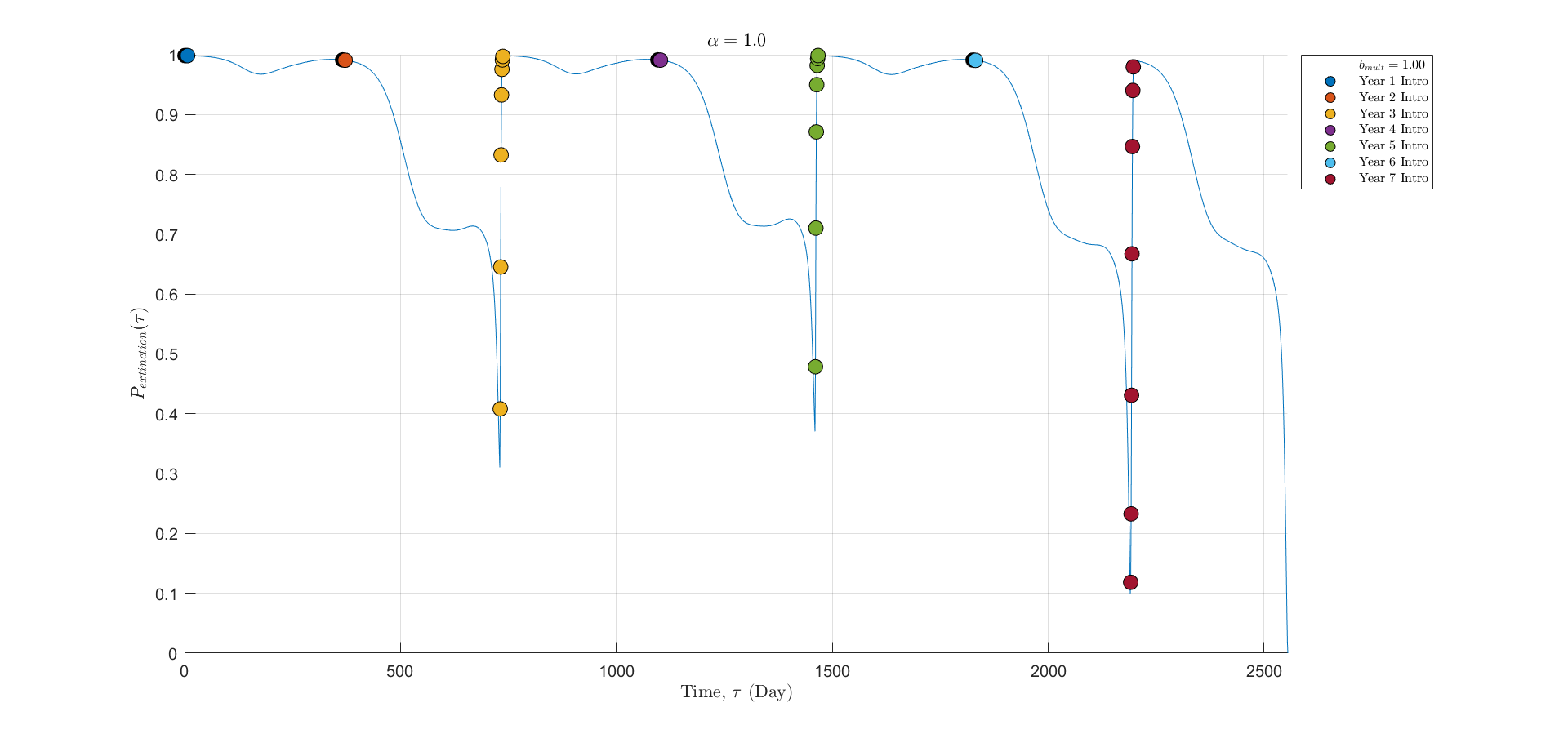}
        \caption{}
        \label{dense_jan}
    \end{subfigure}
    \hspace{.2 cm}
    \begin{subfigure}[c]{0.45\textwidth}
        \centering
        \includegraphics[width=\textwidth]{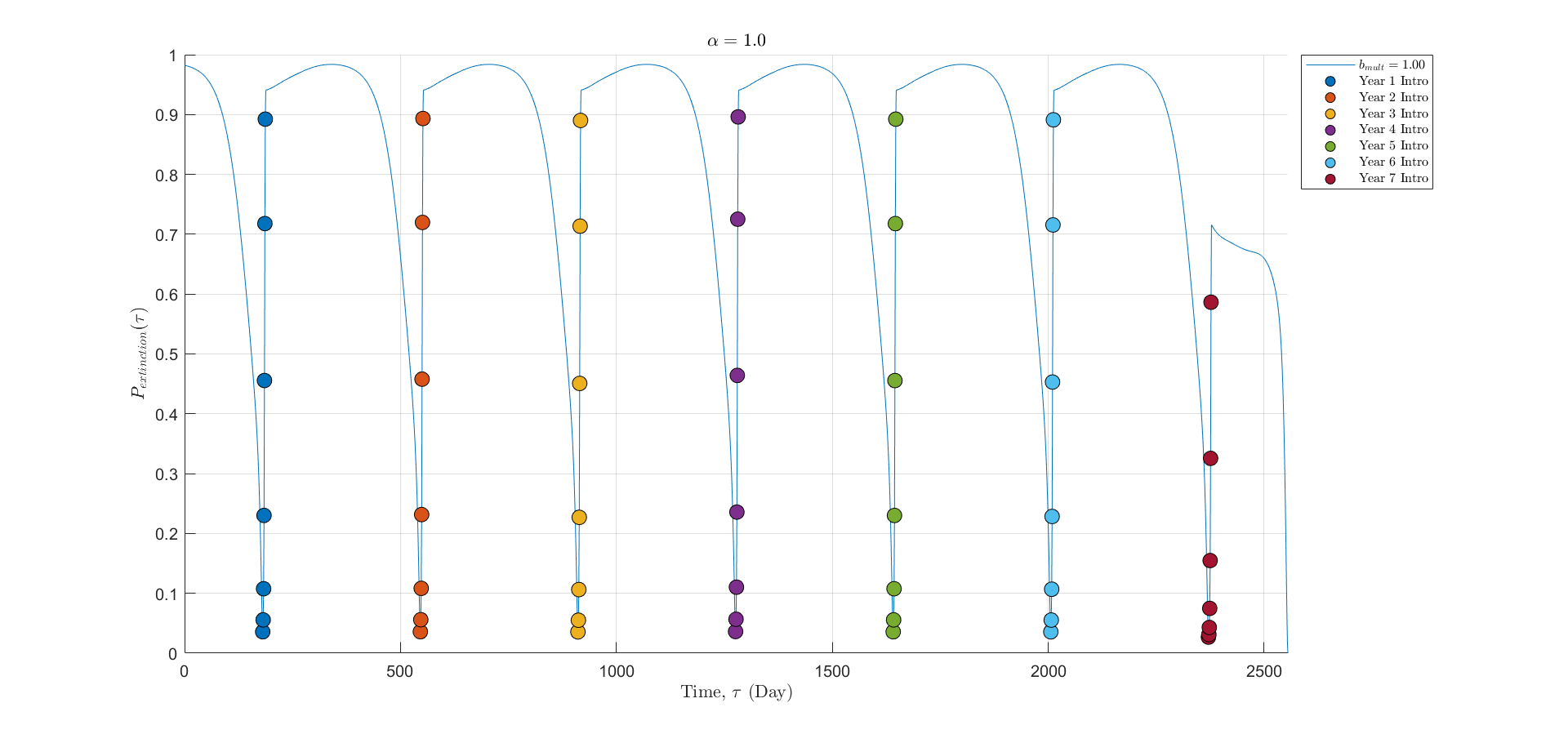}
        \caption{}
        \label{dense_jul}
    \end{subfigure}
       \hspace{.2 cm}
        \begin{subfigure}[c]{0.45\textwidth}
        \centering
        \includegraphics[width=\textwidth]{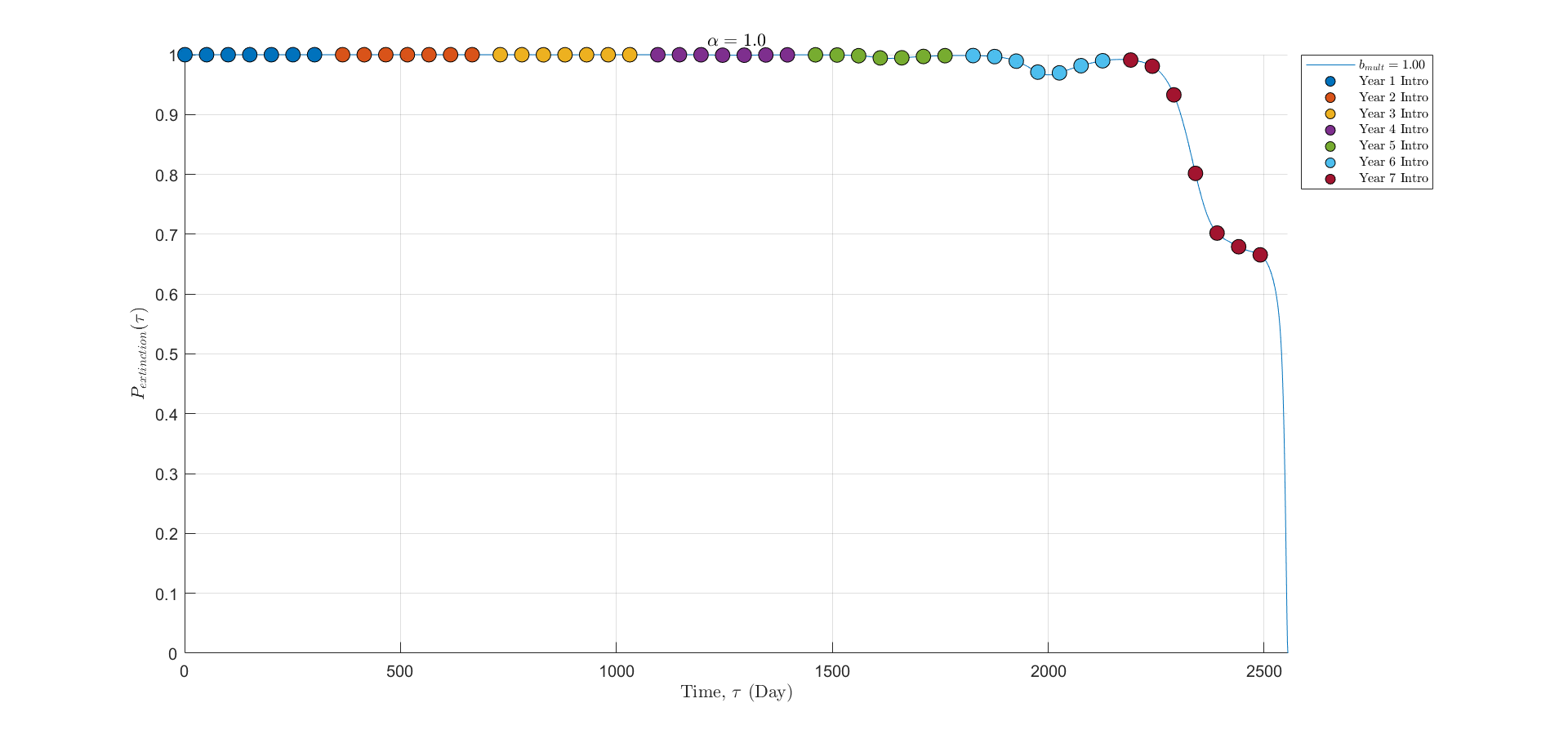}
        \caption{}
        \label{sparsed}
    \end{subfigure}
    
       \caption{Probability of Disease Extinction calculated for yearly periodical densely distributed external introductions of infectious individuals on January $1^{st}$ through January $7^{th}$ in Figure (a), and July $1^{st}$ to July $7^{th}$  in Figure (b). Figure (c) consists of yearly periodical sparsely distributed external introductions of infectious individuals from January to December. Both introductions-sparsely and densely distributed introductions- have the same total number of external infected individuals introduced in Miami-Dade. }
       \label{dense_intro}
\end{figure}

This plot shows the extinction probability $P_i(\tau, t)$, which is the probability that the disease eventually dies out at time $t$, given that a single infectious individual was introduced at some earlier time $\tau < t$. The smooth blue curve represents how extinction probability evolves continuously in time, while the colored dots highlight the specific introduction events that occur periodically across different years. Each cluster of dots corresponds to introductions that happen within the same year. Notice how the extinction probability depends strongly on when the introduction occurs:

If an introduction happens during a period when extinction probability is high (near the peaks of the curve), the chance of fade-out is large. Conversely, if the introduction coincides with a period where the extinction probability dips, i.e, the valleys of the curve, the infection is more likely to persist and spread.

Even though at each $0\leq\tau<t$ we are introducing just one individual for the first time, the periodic pattern of external introductions influences the shape of $P_i(\tau,t)$. This is why the extinction probability oscillates across the years — the timing of introductions interacts with the underlying seasonal dynamics of transmission. The dots, scattered at fixed introduction times each year, show how this periodicity imprints itself year after year: the same “window of vulnerability”, occurring usually during the month of July more intensively, repeats; however the extinction probability shifts depending on how close is $\tau$ to the final time $t>\tau$. In short, the Figure \ref{dense_intro} illustrates that disease extinction is not just a matter of introducing one individual, but of when the introduction occurs relative to the system’s periodic forcing.

\begin{figure}[htbp]
    \centering
    \begin{subfigure}[c]{0.45\textwidth}
        \centering
        \includegraphics[width=\textwidth]{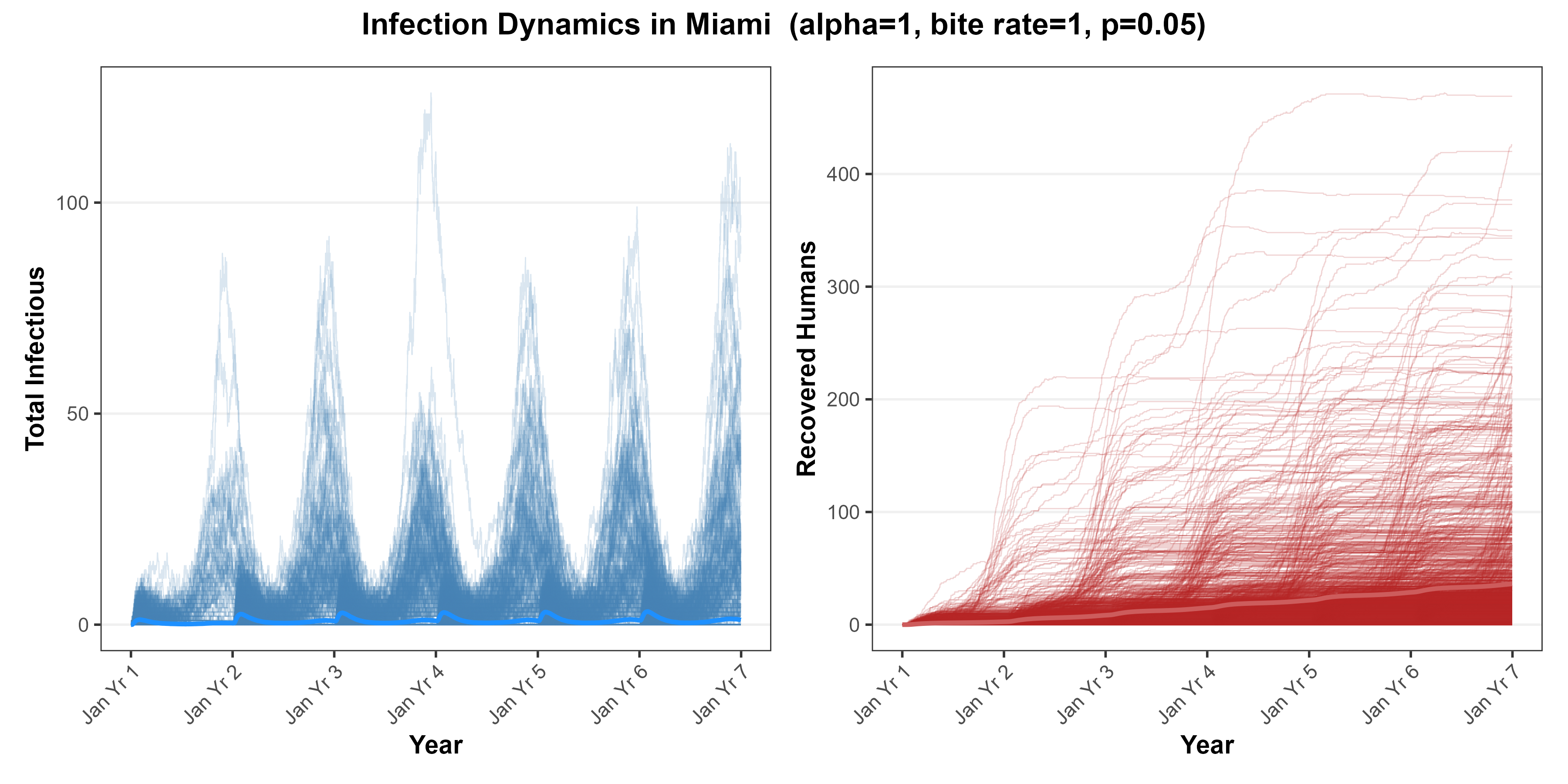}
        \caption{ }
        \label{multi_intr_dense_jan}
    \end{subfigure}
    \hfill
    \begin{subfigure}[c]{0.45\textwidth}
        \centering
        \includegraphics[width=\textwidth]{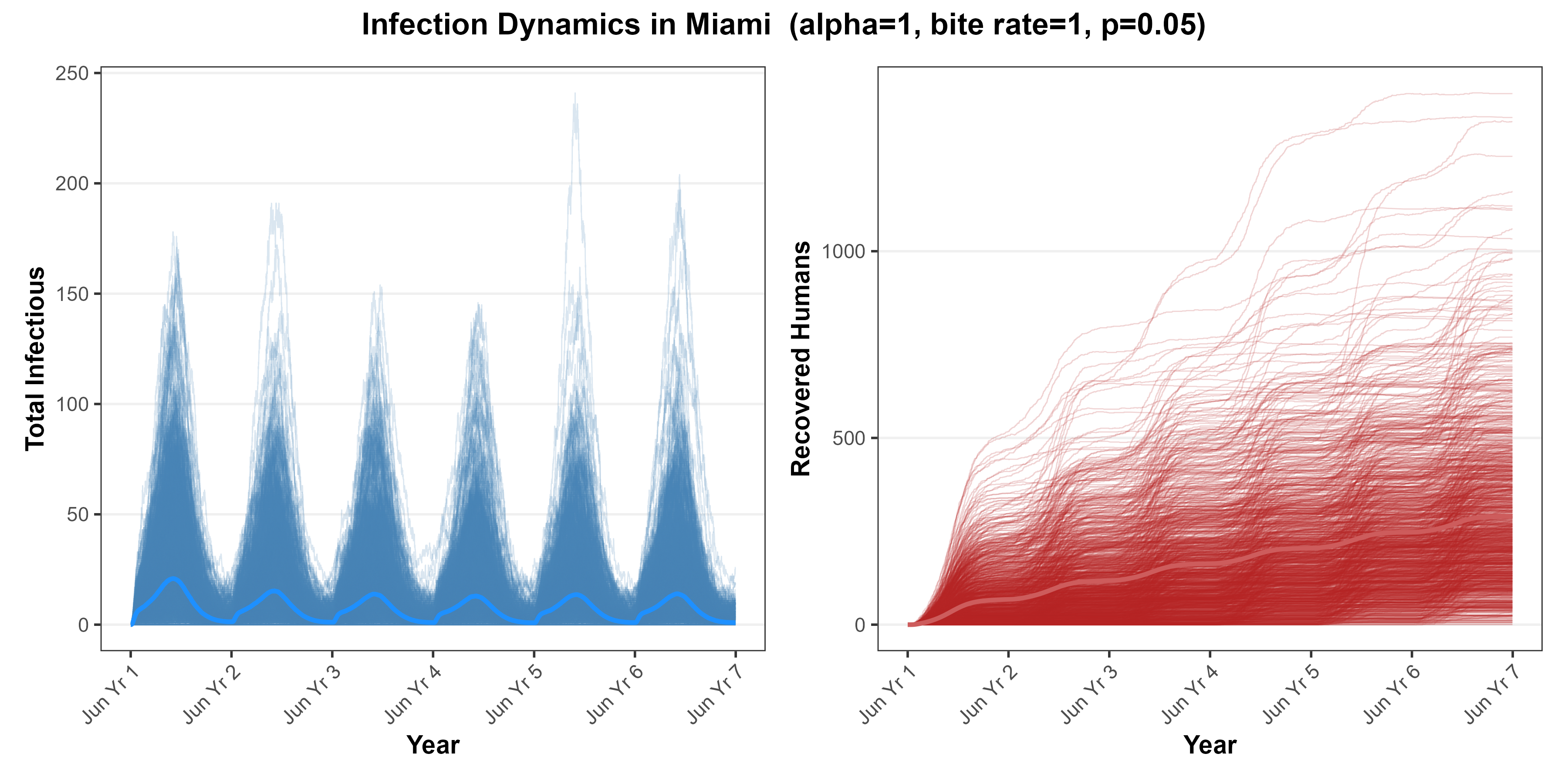}
        \caption{}
        \label{multi_intr_dense_jun}
    \end{subfigure}
    \vspace{0.05cm}
    \begin{subfigure}[b]{0.45\textwidth}
        \centering
        \includegraphics[width=\textwidth]{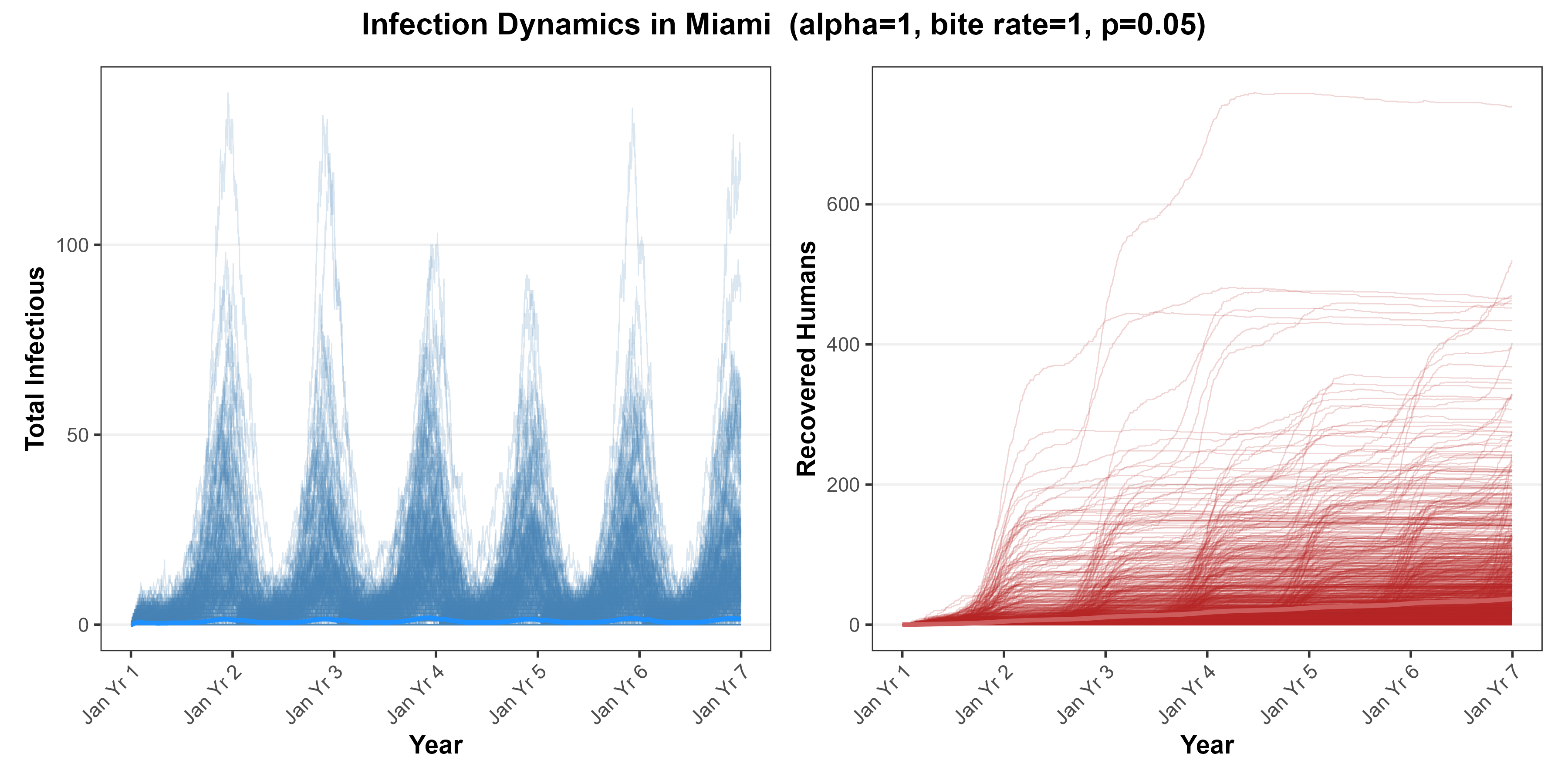}
        \caption{}
        \label{multiple_intro_spaced}
    \end{subfigure}
    \caption{Infectious dynamics for multiple periodical introductions along seven years. Figure (a) refers to densely distributed introductions from January $1^{st}$ to January $7^{th}$ of each year, while Figure (b) refers to the introductions of infectious individuals from July $1^{st}$ to July $7^{th}$ and Figure (c) spaced introduction of seven individuals January through July. In total, 2000 trajectories are plotted and for each figure the solid lines represent the mean.}
    \label{lastfig}
\end{figure}

Figure \ref{lastfig} shows infectious dynamics over 7 years for sparsely and densely distributed introductions. Dense introductions are further divided into those occurring throughout January and July. Dense introductions during summer- e.g, July- result in larger outbreaks and sustained disease persistence across years, highlighting the season's heightened transmission risk.

%DEVO DEIXAR CLARO Q P TODO $\tau>0$, AO CALCULARMOS $P_i(\tau,t)$, assumimos q nenhum invidivuo infeccioso fora introduzido na sociedade ainda, e q estamos introduzindo um unico individuo pela primeira vez neste tempo $\tau>=0$, dado q ha introducoes periodicas de individuos vindos de um compartimento externo. TODAVIA, COMO AS INTRODUCOES SAO PERIODICAS, A FUNCAO DE INTRODUCAO TERA INFLUENCIA NA PROBABILIDADE DE EXTINCAO PARA CADA $\tau$, MESMO Q ATE O EXATO INSTANTE $\tau$ N HAJA UMA INTRODUCAO EXTERNA.

\section{Discussion}

In this study, we develop a Stochastic model that aims to analyze the disease dynamics and to quantify the risks of disease persistence in Miami-Dade, with a particular focus on the effects of seasonal forcing and repeated external introductions of infectious individual. Through the Branching Process Approximation, our approach enables the computation of the probability of disease extinction within 7 years in Miami-Dade, with and without periodical introductions of external infected individuals. Also, by the CTMC models defined in Section 3 and in Section 4, the Chikungunya transmission dynamics was analyzed under different scenarios aiming to consider potential public health interventions targeting to reduce the biting rates and the carrying capacity, as well as considering distinct external introductions of infected individuals during distinct periods of the year. Also, for the case of non-periodical external introductions, we tested the accuracy of the Branching Process Approximation which led to the computation of the probability of disease extinction, according to Figure \ref{test_prob_jan} and Figure \ref{test_prob_june}. It turned out that Assumptions (i)-(iii) were proven to be an accurate approximation to the CTMC model, defined in Table \ref{tab:transition}.

The analysis further explores how the timing and frequency of introductions influence the risk of outbreak. Although single introductions provide information on the baseline capacity of the system to sustain transmission, multiple introductions reveal how clustered or dispersed seeding events alter the likelihood of disease persistence. This dual perspective is essential in settings like Miami-Dade, where periodic importation of cases from endemic regions plays a central role in the dynamics of local outbreaks. As shown in Figure \ref{dense_intro} and Figure \ref{lastfig}, sparsely distributed introductions or densely distributed introductions during the Winter, e.g multiple introductions during consecutive days in January, induce a significantly greater probability of disease extinction within 7 years, compared to densely distributed introduction during the Summer, e.g, multiple introductions during consecutive days in July.

\bibliographystyle{plain}  % You can use "abbrv", "unsrt", etc.
\bibliography{references}  % Make sure the file is named references.bib

\end{document}